\documentclass[letterpaper,twocolumn,10pt]{article}
\usepackage{usenix-2020-09}

\usepackage{tikz}
\usepackage{amsmath,amssymb}

\usepackage{filecontents}

\usepackage{nicefrac}
\usepackage{siunitx}
\usepackage{array,framed}
\usepackage{booktabs}
\usepackage{
  color,
  float,
  epsfig,
  wrapfig,
  graphics,
  graphicx,
  subcaption
}
\usepackage{setspace}
\usepackage{latexsym,fancyhdr,url}
\usepackage{enumerate}
\usepackage[algo2e]{algorithm2e}
\usepackage{algorithmic}
\usepackage{algorithm}
\usepackage{graphics}
\usepackage{xparse} 
\usepackage{xspace}
\usepackage{multirow}
\usepackage{csvsimple}
\usepackage{balance}
\usepackage{mathrsfs}
\usepackage{amsfonts}
\usepackage{epsfig}
\usepackage{pifont}

\let\labelindent\relax
\usepackage{enumitem}
\usepackage{mathtools}
\usepackage{tikz}
\usepackage{pgfplots} 
\usepackage{pgfplotstable}
\pgfplotsset{compat = newest}
\usetikzlibrary{arrows}
\usetikzlibrary{patterns}
\usepackage{stackrel}
\usepackage{pstricks}
\usepackage{subcaption}
\usepackage{wrapfig}
\usepackage[utf8]{inputenc}
\usepackage{booktabs}
\usepackage{expl3}
\usepackage{collcell}
\usepackage{colortbl} 
\usepackage{xurl}
\usepackage{hyperref}
\usepackage{rotating}
\usetikzlibrary{
  shapes.geometric,
  arrows,
  external,
  pgfplots.groupplots,
  matrix
}

\DeclareMathAlphabet{\mathcal}{OMS}{cmsy}{m}{n}

\DeclareGraphicsExtensions{%
    .png,.PNG,%
    .pdf,.PDF,%
    .jpg,.mps,.jpeg,.jbig2,.jb2,.JPG,.JPEG,.JBIG2,.JB2}

\newcommand{\sysName}{Mendata\xspace}
\newcommand{\myparagraph}[1]{\smallskip\noindent\textbf{#1}:\xspace}

\newcounter{note}[section]

\newcommand{\notecolor}{blue}
\renewcommand{\thenote}{\thesection.\arabic{note}}
\newcommand{\mkr}[1]{\refstepcounter{note}{\bf \textcolor{\notecolor}{$\ll$MKR~\thenote: {\sf #1}$\gg$}}}
\newcommand{\zh}[1]{\refstepcounter{note}{\bf \textcolor{red}{$\ll$ZH~\thenote: {\sf #1}$\gg$}}}

\newcommand{\secref}[1]{\mbox{Sec.~\ref{#1}}\xspace}
\newcommand{\Lineref}[1]{\mbox{Line~\ref{#1}}\xspace}

\newcommand{\linesref}[2]{\mbox{Lines~\ref{#1}--\ref{#2}}\xspace}

\newcommand{\tblref}[1]{\mbox{Table~\ref{#1}}\xspace}

\newcommand{\appref}[1]{\mbox{App.~\ref{#1}}\xspace}
\newcommand{\eqnref}[1]{\mbox{Eq.~(\ref{#1})}\xspace}
\newcommand{\eqnsref}[2]{\mbox{Eqns.~(\ref{#1})--(\ref{#2})}\xspace}

\newcommand{\algref}[1]{\mbox{Alg.~(\ref{#1})}\xspace}

\newcommand{\Norm}[2]{\ensuremath{\left\Vert{#2}\right\Vert_{#1}}\xspace}

\newcommand{\euclideanNorm}[1]{\ensuremath{\Norm{2}{#1}}\xspace}
\newcommand{\infinityNorm}[1]{\ensuremath{\Norm{\infty}{#1}}\xspace}

\NewDocumentCommand{\genericNat}{ g }{\ensuremath{z\IfNoValueF{#1}{_{#1}}}\xspace}
\newcommand{\genericVec}{\ensuremath{v}\xspace}
\newcommand{\genericVecAlt}{\ensuremath{v'}\xspace}
\newcommand{\setSize}[1]{\ensuremath{\left|{#1}\right|}\xspace}

\newcommand{\reals}{\ensuremath{\mathbb{R}}\xspace}

\newcommand{\floor}[1]{\ensuremath{\lfloor{#1}\rfloor}\xspace}
\NewDocumentCommand{\expv}{ o g }{\ensuremath{\mathbb{E}\IfNoValueF{#1}{_{#1}}\left({#2}\right)}\xspace}
\NewDocumentCommand{\gradient}{ g }{\ensuremath{\nabla\IfNoValueF{#1}{_{#1}}}\xspace}
\newcommand{\cset}[3]{\ensuremath{#1\{}{#2}\ensuremath{\;#1|} \ifmmode{\;}\fi {#3}\ensuremath{#1\}}\xspace}
\newcommand{\cprob}[3]{\ensuremath{\mathbb{P}#1(}{#2}\ensuremath{\;#1|} \ifmmode{\;}\fi {#3}\ensuremath{#1)}\xspace}
\newcommand{\cexpv}[3]{\ensuremath{\mathbb{E}#1(}{#2}\ensuremath{\;#1|} \ifmmode{\;}\fi {#3}\ensuremath{#1)}\xspace}

\NewDocumentCommand{\Dataset}{ g g }{\ensuremath{\mathrm{D}\IfNoValueF{#1}{^{#1}}\IfNoValueF{#2}{_{#2}}}\xspace}
\NewDocumentCommand{\Image}{ g g }{\ensuremath{x\IfNoValueF{#1}{^{#1}}\IfNoValueF{#2}{_{#2}}}\xspace}

\newcommand{\NumClasses}{\ensuremath{\mathrm{Y}}\xspace}
\newcommand{\ClassesSpace}{\ensuremath{\mathcal{Y}}\xspace}

\NewDocumentCommand{\Accuracy}{g}{\ensuremath{\mathsf{acc}\IfNoValueF{#1}{_{#1}}}\xspace}
\NewDocumentCommand{\ManipulateFunction}{g g}{\ensuremath{\textsc{Man}\IfNoValueF{#1}{_{#1}}\IfNoValueF{#2}{^{#2}}}\xspace}
\NewDocumentCommand{\MarkFunction}{ g g }{\ensuremath{\textsc{Mar}\IfNoValueF{#1}{^{#1}}\IfNoValueF{#2}{_{#2}}}\xspace}
\NewDocumentCommand{\VerifyFunction}{ g g}{\ensuremath{\textsc{Ver}\IfNoValueF{#1}{^{#1}}\IfNoValueF{#2}{_{#2}}}\xspace}
\NewDocumentCommand{\PurifyFunction}{ g }{\ensuremath{\textsc{Pur}\IfNoValueF{#1}{^{#1}}}\xspace}

\newcommand{\trainTag}{\ensuremath{\mathsf{tr}}\xspace}

\newcommand{\markedTag}{\ensuremath{\mathsf{mk}}\xspace}
\newcommand{\referenceTag}{\ensuremath{\mathsf{rf}}\xspace}
\newcommand{\purifiedTag}{\ensuremath{\mathsf{pf}}\xspace}

\newcommand{\untrustedTag}{\ensuremath{\mathsf{ut}}\xspace}
\newcommand{\perturbedTag}{\ensuremath{\mathsf{pe}}\xspace}

\NewDocumentCommand{\Network}{g}{\ensuremath{f\IfNoValueF{#1}{^{#1}}}\xspace}
\newcommand{\LinearClassifier}[1]{\ensuremath{g{#1}}\xspace}
\NewDocumentCommand{\NetworkParams}{ g g}{\ensuremath{\theta\IfNoValueF{#1}{^{#1}}\IfNoValueF{#2}{_{#2}}}\xspace}

\newcommand{\noise}{\ensuremath{\xi}\xspace}

\newcommand{\KLDivergence}[2]{\ensuremath{D_{KL}\big({#1}\parallel{#2}\big)}\xspace}
\NewDocumentCommand{\ReferenceMethod}{ g }{\ensuremath{\textsc{Ref}\IfNoValueF{#1}{(#1)}}\xspace}
\NewDocumentCommand{\RandomNoise}{ g }{\ensuremath{\textsc{Rand}\IfNoValueF{#1}{(#1)}}\xspace}
\NewDocumentCommand{\RandomNoiseAlt}{ g }{\ensuremath{\textsc{Rand*}\IfNoValueF{#1}{(#1)}}\xspace}
\NewDocumentCommand{\FriendlyNoise}{ g }{\ensuremath{\textsc{Frien}\IfNoValueF{#1}{(#1)}}\xspace}
\NewDocumentCommand{\FriendlyNoiseAlt}{ g }{\ensuremath{\textsc{Frien*}\IfNoValueF{#1}{(#1)}}\xspace}
\NewDocumentCommand{\FriendlyNoiseB}{ g }{\ensuremath{\textsc{FrienB}\IfNoValueF{#1}{(#1)}}\xspace}
\NewDocumentCommand{\EPIC}{ g }{\ensuremath{\textsc{Epic}\IfNoValueF{#1}{(#1)}}\xspace}
\NewDocumentCommand{\ASD}{ g }{\ensuremath{\textsc{ASD}\IfNoValueF{#1}{(#1)}}\xspace}
\NewDocumentCommand{\CBD}{ g }{\ensuremath{\textsc{CBD}\IfNoValueF{#1}{(#1)}}\xspace}

\newcommand{\BernoulliNoiseParameter}{\ensuremath{\varkappa}\xspace}
\newcommand{\PerturbationMagnitude}{\ensuremath{\alpha}\xspace}
\newcommand{\GassianVariant}{\ensuremath{\sigma}\xspace}
\newcommand{\FriendlynoiseHyperparameter}{\ensuremath{\mu}\xspace}
\newcommand{\EPICEpochParameter}{\ensuremath{\digamma}\xspace}
\newcommand{\EPICWarmupParameter}{\ensuremath{\varpi}\xspace}
\newcommand{\EPICSubsetSizeParameter}{\ensuremath{\varsigma}\xspace}
\NewDocumentCommand{\ASDEpochParameter}{ g }{\ensuremath{\digamma\IfNoValueF{#1}{_{#1}}}\xspace}

\newcommand{\classIdx}{\ensuremath{i}\xspace}

\newcommand{\DistributionDistance}{\ensuremath{\mathsf{d}}\xspace}
\newcommand{\WassersteinDistance}{\ensuremath{\mathsf{d}}\xspace}
\NewDocumentCommand{\Critic}{g}{\ensuremath{h\IfNoValueF{#1}{_{#1}}}\xspace}
\NewDocumentCommand{\CriticSpace}{g}{\ensuremath{H\IfNoValueF{#1}{_{#1}}}\xspace}
\newcommand{\Distribution}[1]{\ensuremath{\mathbb{P}^{#1}}\xspace}

\newcommand{\NumSamples}{\ensuremath{n}\xspace}
\newcommand{\SampleDimension}{\ensuremath{w}\xspace}
\newcommand{\CheckingPoint}{\ensuremath{t}\xspace}
\newcommand{\Regularization}{\ensuremath{r}\xspace}

\newcommand{\ImageRange}{\ensuremath{[0,1]}\xspace}
\newcommand{\MetricSet}{\ensuremath{\mathcal{V}}\xspace}

\newcommand{\AveragedFunction}{\ensuremath{\mathsf{Avg}}\xspace}
\NewDocumentCommand{\FeatureVector}{g g}{\ensuremath{v\IfNoValueF{#1}{_{#1}}\IfNoValueF{#2}{^{#2}}}\xspace}
\newcommand{\JointDistributionSet}{\ensuremath{\Pi}\xspace}
\newcommand{\JointDistribution}{\ensuremath{\gamma}\xspace}
\NewDocumentCommand{\PurifiedPerturbation}{g}{\ensuremath{\delta\IfNoValueF{#1}{_{#1}}}\xspace}

\newcommand{\PurificationHyperparameter}{\ensuremath{\lambda}\xspace}
\newcommand{\BatchSize}{\ensuremath{m}\xspace}
\newcommand{\IdxInBatch}{\ensuremath{b}\xspace}
\newcommand{\DataBatch}[2]{\ensuremath{B_{#1}^{#2}}\xspace}
\newcommand{\LearningRate}[1]{\ensuremath{l_{#1}}\xspace}
\newcommand{\AdamOptimizer}{\ensuremath{\mathsf{Adam}}\xspace}

\NewDocumentCommand{\SampleIdx}{g}{\ensuremath{j\IfNoValueF{#1}{_{#1}}}\xspace}
\NewDocumentCommand{\SampleIdxAlt}{g}{\ensuremath{z\IfNoValueF{#1}{_{#1}}}\xspace}
\newcommand{\RandomNumber}{\ensuremath{\eta}\xspace}
\NewDocumentCommand{\LossFunction}{ g g}{\ensuremath{\ell\IfNoValueF{#1}{_{#1}}\IfNoValueF{#2}{^{#2}}}\xspace}
\newcommand{\Gradient}[1]{\ensuremath{\nabla_{#1}}\xspace}
\newcommand{\SelectionRatio}{\ensuremath{\rho}\xspace}
\newcommand{\GradientPenaltyCoefficient}{\ensuremath{\nu}\xspace}
\NewDocumentCommand{\CriticParams}{ g }{\ensuremath{\vartheta\IfNoValueF{#1}{_{#1}}}\xspace}
\newcommand{\Amplification}{\ensuremath{\beta}\xspace}

\newcommand{\AttackSuccessRate}{\ensuremath{\mathsf{ASR}}\xspace}
\NewDocumentCommand{\ModelParams}{ g }{\ensuremath{w\IfNoValueF{#1}{_{#1}}}\xspace}
\newcommand{\PoisonHyperparameter}{\ensuremath{\epsilon}\xspace}
\newcommand{\TracingHyperparameter}{\ensuremath{\epsilon}\xspace}

\newcommand{\Precision}{\ensuremath{q}\xspace}
\newcommand{\Recall}{\ensuremath{r}\xspace}

\NewDocumentCommand{\Label}{ g g }{\ensuremath{y\IfNoValueF{#1}{^{#1}}\IfNoValueF{#2}{_{#2}}}\xspace}
\newcommand{\knnParameter}{\ensuremath{k}\xspace}

\newcommand{\NumThreshold}{\ensuremath{\kappa}\xspace}
\NewDocumentCommand{\ManipulationPerturbation}{ g }{\ensuremath{\zeta\IfNoValueF{#1}{_{#1}}}\xspace}

\newcommand{\TargetImage}{\ensuremath{x'}\xspace}
\newcommand{\TargetLabel}{\ensuremath{y'}\xspace}
\newcommand{\Trigger}{\ensuremath{z}\xspace}

\newcommand{\Mark}[1]{\ensuremath{u_{#1}}\xspace}

\newcommand{\PValue}{\ensuremath{p}\xspace}

\ExplSyntaxOn
\cs_new:Npn \intensity #1
   {\fp_eval:n {1.0 - (\MinIntensity + (1.0-\MinIntensity) * ((({#1}-\value{MinNumber})/(\value{MaxNumber}-\value{MinNumber}))))}
   }
\ExplSyntaxOff
\newcommand{\MinIntensity}{0.0}   
\newcounter{MinNumber}
\newcounter{MaxNumber}

\newcommand{\ApplyGradientX}[1]{\cellcolor[gray]{\intensity{#1}}}
\newcolumntype{X}{>{\collectcell\ApplyGradientX}p{1.25em}<{\endcollectcell}}

\begin{document}
%
\title{\sysName: A Framework to Purify Manipulated Training Data}

\author{
{\rm Zonghao Huang}\\
Duke University
\and
{\rm Neil Gong}\\
Duke University
\and
{\rm Michael K. Reiter}\\
Duke University
} 

\maketitle


\begin{abstract}
  Untrusted data used to train a model might have been manipulated to
  endow the learned model with hidden properties that the data
  contributor might later exploit.  \textit{Data purification} aims to
  remove such manipulations prior to training the model.  We propose
  \sysName, a novel framework to purify manipulated training data.
  Starting from a small \emph{reference dataset} in which a large
  majority of the inputs are clean, \sysName perturbs the training
  inputs so that they retain their utility but are distributed
  similarly (as measured by Wasserstein distance) to the reference
  data, thereby eliminating hidden properties from the learned model.
  A key challenge is how to find such perturbations, which we address
  by formulating a min-max optimization problem and developing a
  two-step method to iteratively solve it.  We demonstrate the
  effectiveness of \sysName by applying it to defeat state-of-the-art
  data poisoning and data tracing techniques.

\end{abstract}

\section{Introduction} \label{sec:introduction}

A machine learning (ML) system often harvests training data from
unverified \emph{data contributors}, e.g., users/webpages on the
Internet. In such cases, an ML system is fundamentally vulnerable to
\emph{data manipulation}, i.e., changes introduced to the training
data to endow the learned model with certain hidden,
data-contributor-desired \emph{properties}. Such data manipulation
techniques can be used as an attack to compromise the integrity of an
ML system or as a defense to protect data contributors.  For instance,
when a malicious data contributor \textit{poisons} the training
data~\cite{shafahi2018:poison, zhu2019:transferable,
  huang2020:metapoison, geiping2020:witches, schwarzschild2021:just},
the property is that the learned model will misclassify (perhaps in a
targeted way) a certain \textit{target input} or any input patched
with a \textit{trigger}, pre-chosen by the data contributor. When a
data contributor instead marks its data for the purpose of
\textit{data tracing}~\cite{sablayrolles2020:radioactive}, the
property is that the data contributor can detect that her data was
used for model training (perhaps without authorization).

Most existing defenses against data manipulation focus on
\emph{prevention}~\cite{lecuyer2019:certified,rosenfeld2020:certified,zhang2022:bagflip,ma2019:data,hong2020:effectiveness,jia2021:intrinsic,levine2021:deep,jia2022:certified,yang2022:not} or \emph{detection}~\cite{peri2020:deep,chen2018:detecting,tran2018:spectral}. Specifically, a
prevention mechanism aims to redesign the ML training algorithm so
that the learned model does not inherit the hidden property even if
some training inputs are manipulated. For instance, ensemble methods,
which train multiple models on different subsets of the training data,
can guarantee that the target input is correctly classified, despite
some training data having been poisoned~\cite{jia2021:intrinsic,levine2021:deep,jia2022:certified}. However, these
prevention methods 1) require modifying the ML training algorithm, 2)
can only tolerate a tiny fraction of poisoned training inputs, and 3)
often sacrifice the utility of the models. A detection mechanism aims
to detect clean training data in a manipulated training dataset, and
only the training data detected to be clean is subsequently used for
model training. These detection methods can effectively make the
learned model free of hidden properties, but they sacrifice utility of
the learned model by discarding the training data that are (perhaps
falsely) detected as manipulated.

\emph{Data purification} aims to complement prevention and
detection. For instance, when a subset of the manipulated training
dataset is detected as clean by a detector, instead of throwing away
the remaining training data, data purification can be used to purify
them.  We refer to the training data detected to be clean as
\emph{reference data} and the remaining training data as
\emph{untrusted data}. Then, the \emph{purified training data}, which
consists of the reference data and the purified untrusted data, is
used for model training. Note that the reference dataset may include
poisoned training data while the untrusted dataset may include clean
training data, due to the imperfection of the detector.
Unfortunately, data purification is largely unexplored, and indeed, we
are aware of only one instance: Liu et al.~\cite{liu2022:friendly} proposed a data
purification method called \emph{Friendly Noise}, which adds as large
perturbations to the untrusted dataset as possible while maintaining
the predictions made by a model pre-trained on the original
manipulated training dataset.

We propose \emph{\sysName}, a novel and general framework to purify
(or ``mend'') manipulated training data. Given a manipulated training
dataset, a subset of which is detected as a reference dataset by an
existing detector, \sysName aims to purify the inputs in the remaining
untrusted dataset by adding carefully crafted perturbations to
them. Specifically, \sysName aims to achieve three goals: 1) a
\emph{property goal}, which is that a model learned on the purified
training data is free of the data-contributor-desired property, 2) a
\emph{utility goal}, which is that a model learned on the purified
training data is accurate, and 3) a \emph{generality goal}, which is
that \sysName is applicable to a wide range of data manipulation
algorithms and properties.

To achieve the property goal, our key observation is that a model
trained on the reference dataset is unlikely to have the property
since an overwhelming majority of the reference data are clean
(assuming the detector is reasonably accurate).  Therefore, \sysName
perturbs each input in the untrusted dataset so that they follow a
similar distribution to the reference dataset.  We quantify the
distribution similarity between the two datasets using the well-known
\emph{Wasserstein distance}~\cite{vaserstein1969:markov,gulrajani2017:improved}, due to its intuitive
interpretation (the minimum cost of moving a pile of earth in the
shape of one distribution to achieve the shape of the other). To
achieve the utility goal, \sysName aims to bound the perturbations
added to the untrusted inputs. Formally, we formulate finding the
perturbations as a minimization optimization problem, whose objective
function is a weighted sum of the Wasserstein distance between the two
dataset distributions (quantifying the property goal) and the
magnitudes of the perturbations (quantifying the utility goal).
\sysName achieves the generality goal by not depending on any specific
data manipulation algorithm or property it might be intended to
embed in the learned model, when formulating and solving the
optimization problem.

However, it is hard to solve the optimization problem due to the
challenge of computing the Wasserstein distance. To address the
challenge, we leverage the Kantorovich-Rubinstein duality~\cite{villani2009:optimal} to
approximate the Wasserstein distance as the \emph{maximum} difference
between the average function value of the untrusted inputs and that of
the reference inputs, where the function value of an input is
calculated using a neural network that implements a $1$-Lipschitz
function. Then, we refine our optimization problem as a min-max one,
where the inner max problem aims to find the $1$-Lipschitz neural
network that approximates the Wasserstein distance and the outer min
problem aims to find the perturbations.  We detail a two-step method
to iteratively solve the min-max optimization problem.  One step
starts from current perturbations and solves the inner max problem to
update the neural network used to approximate the Wasserstein
distance.  The second step starts from the current neural network
approximating Wasserstein distance and solves the outer min problem to
update the perturbations.


We evaluate \sysName on two representative use cases. The first use
case is to mitigate data poisoning
attacks~\cite{huang2020:metapoison,geiping2020:witches,souri2022:sleeper},
while the second is to avoid data
tracing~\cite{sablayrolles2020:radioactive}. Our results on multiple
image benchmark datasets show that \sysName can effectively eliminate
the hidden property from the model learned on the purified training
data while still achieving high accuracy, even if the reference
dataset includes a small fraction of poisoned data. Moreover, we show
that \sysName outperforms existing data purification methods, notably
Friendly Noise~\cite{liu2022:friendly}, and state-of-the-art
prevention methods.  For instance, when the models learned on the
purified training data achieve similar accuracy, the models learned on
the training data purified by \sysName are more likely to be
property-free.

To summarize, our key contributions are as follows:
\begin{itemize}[nosep,leftmargin=1em,labelwidth=*,align=left]
\item We propose \sysName, a general framework to purify manipulated
  training data.

\item We formulate data purification as an optimization problem and
  leverage a two-step method to approximately solve it.

\item We show the effectiveness of \sysName by applying it to two
  representative use cases, namely to defeat state-of-the-art data
  poisoning and data tracing schemes.
\end{itemize}


\section{Related Work} \label{sec:related}

\subsection{Training-Data Manipulation} \label{sec:related:manipulation}

\myparagraph{Clean-label vs. dirty-label training-data manipulation}
Training-data manipulation generally refers to tampering with the
training data to endow the learned model with certain hidden
properties. Some training-data manipulation methods~\cite{huang2020:metapoison,geiping2020:witches,aghakhani2021:bullseye,schwarzschild2021:just,sablayrolles2020:radioactive} only
tamper with the training inputs (known as \emph{clean-label}), while
some~\cite{biggio2011:support,liu2018:trojaning,wenger2021:backdoor} also tamper with the labels (known as \emph{dirty-label}). For
instance, dirty-label backdoor attacks~\cite{chen2017:targeted,gu2019:badnets}, which embed an attacker-chosen
trigger into some training inputs and re-label them as an
attacker-chosen, incorrect label, are representative dirty-label
training-data manipulation methods. However, labels are often manually
analyzed and assigned by a data curator depending on the application
context of the data. For instance, a face image could be labeled as
the person name when used for face recognition~\cite{zhao2003:face}, or labeled as
``happy'' when used for emotion recognition~\cite{han2014:speech}. In such cases, it is hard
for a data contributor to tamper with the labels. Therefore, we focus
on \emph{clean-label} training-data manipulation in this work.

\myparagraph{Training-data manipulation as attacks} Training data
manipulation can be used to compromise the integrity of an ML
system. Such attacks are known as \textit{data poisoning attacks} and
can be categorized into \textit{untargeted attacks}~\cite{biggio2012:poisoning,xiao2015:feature,jagielski2018:manipulating},
\textit{subpopulation attacks}~\cite{jagielski2021:subpopulation},
\textit{targeted attacks}~\cite{shafahi2018:poison,zhu2019:transferable,huang2020:metapoison,geiping2020:witches,yang2022:not},
and \textit{backdoor attacks}~\cite{turner2018:clean,zhao2020:clean,souri2022:sleeper}. Untargeted
attacks decrease the testing accuracy of the learned model for many
inputs, while subpopulation (or targeted) attacks cause the learned
model to misclassify an attacker-chosen group of inputs (or a single
\emph{target input}) as an attacker-chosen \emph{target label}.
Clean-label backdoor attacks mislead the trained model to classify inputs
patched with an attack-chosen \emph{trigger} as a \emph{target label} by inserting perturbations
into the training data without necessarily changing their labels.
Targeted attacks (or backdoor attacks) do not affect the testing accuracy of the learned
model for the non-target inputs (or normal inputs), which makes it harder to be detected
than the other two categories of attacks. We show in
\secref{sec:poisoning} that \sysName can be applied to defend against
targeted attacks and backdoor attacks.

\myparagraph{Training-data manipulation as defenses} Training data
manipulation can also be used as defenses to protect data
contributors~\cite{shan2020:fawkes,sablayrolles2020:radioactive,li2022:untargeted}.  \emph{Data tracing}~\cite{sablayrolles2020:radioactive}, which proactively detects
(unauthorized) data use in model training, is one such representative
use case. In particular, a data contributor carefully
marks/manipulates her data so that a model learned using her marked
data (perhaps without her authorization) can be detected as such. For
instance, \textit{Radioactive Data}~\cite{sablayrolles2020:radioactive}, the state-of-the-art data
tracing method, carefully marks a subset of the training inputs such
that the parameters of a model learned on the marked training data
have a certain bias, which can be detected by hypothesis testing with
statistical guarantees. In \secref{sec:tracing}, we apply our \sysName
as an attack to defeat data tracing.

\subsection{Countering Training-Data Manipulation} \label{sec:related:defense}
\myparagraph{Prevention} A prevention countermeasure against
training-data manipulation aims to redesign the ML training algorithm
so that the learned model is guaranteed to be free of the hidden
properties even if some training examples are arbitrarily
manipulated. For instance, the learned model is guaranteed to not
misclassify the target input as the target label to prevent targeted
data poisoning attacks.  Examples of prevention methods include
\textit{randomized smoothing}~\cite{lecuyer2019:certified,rosenfeld2020:certified,zhang2022:bagflip}, \textit{differential
privacy}~\cite{dwork2006:differential,ma2019:data,hong2020:effectiveness}, 
\textit{majority vote}~\cite{jia2021:intrinsic,levine2021:deep,jia2022:certified},
\textit{and reweighting}~\cite{li2021:anti,yang2022:not,huang2022:backdoor,zhang2023:backdoor,gao2023:backdoor}. Randomized
smoothing adds random noise to the training inputs when learning a
model; differential privacy adds random noise to the gradients used to
update the model during training; and majority vote trains multiple
models on different subsets of the training data~\cite{jia2021:intrinsic,levine2021:deep} or leverages
the intrinsic majority vote mechanism in the nearest neighbor
algorithm~\cite{jia2022:certified}. However, these prevention methods require modifying
the ML training algorithm and often sacrifice model accuracy. 
Reweighting methods dynamically reweight training samples during training but most of them~\cite{li2021:anti,huang2022:backdoor,zhang2023:backdoor,gao2023:backdoor} are 
only effective against dirty-label backdoor attacks~\cite{gu2019:badnets,liu2018:trojaning,nguyen2020:input} and some simple
clean-label backdoor attacks~\cite{turner2018:clean,barni2019:new}.

\myparagraph{Detection} A detection mechanism~\cite{peri2020:deep,zeng2023:meta} aims to detect the clean
training data in the manipulated dataset and only use those data for
model training. For instance, Peri et al.~\cite{peri2020:deep} proposed a
\knnParameter-Nearest Neighbors (\knnParameter-NN) based method,
which is state-of-the-art detection method against clean-label
training-data manipulation. Specifically, \knnParameter-NN detects a
training example as clean if its label is the same as the majority of
its \knnParameter nearest neighbors.  Other studies~\cite{tran2018:spectral,chen2018:detecting,ma2019:nic,gao2019:strip} proposed
detection methods for dirty-label training-data manipulation such as
backdoor attacks. These detection
methods are not applicable for clean-label training-data manipulation,
which is the focus of this work, because they rely on the incorrect
labels of the manipulated training inputs.

\myparagraph{Purification} Purification aims to purify the training
data detected as manipulated, so the training data detected as clean
by a detector and the purified version of those detected as
manipulated can be used together for model training. Unfortunately,
purifying manipulated training data is largely unexplored. To the best
of our knowledge, \textit{Friendly Noise}~\cite{liu2022:friendly} is the only
training-data purification method. Specifically, Friendly Noise aims
to add as large perturbations to the training inputs as possible
without changing their predictions made by a model pre-trained on the
original manipulated training data. However, the model learned on the
original manipulated training data is different from the one learned
on the purified training data. As a result, Friendly Noise achieves a
suboptimal trade-off between model accuracy and effectiveness at
eliminating the hidden properties from the learned model, as shown in
our experimental results.

Some methods~\cite{yoon2021:adversarial,nie2022:diffusion} were proposed to purify \textit{testing} inputs,
e.g., adversarial examples. These methods require a large amount of
\emph{clean} training data to learn generative models (e.g., diffusion
models~\cite{sohl2015:deep}), which are subsequently used to purify testing
inputs. They are not applicable in our scenario where the training
data is manipulated.


 \section{Problem Formulation} \label{sec:problem-statement}

\subsection{System Model} \label{sec:problem-statement:model}

We consider a general machine-learning pipeline with three
entities: a \textit{data contributor}, a \textit{data curator}, and a
\textit{model trainer}.
We describe these three entities as follows.

\myparagraph{Data contributor} A data contributor holds a clean
dataset.  The goal of the data contributor is to turn it into a
manipulated dataset using a \emph{data manipulation algorithm}, such
that a model trained on the manipulated data has a certain hidden
\emph{property}. For instance, in data poisoning attacks~\cite{geiping2020:witches,souri2022:sleeper}, the
property is that the trained model predicts an attacker-chosen label
for an attacker-chosen target input or inputs 
with an attacker-chosen trigger; and in data tracing~\cite{sablayrolles2020:radioactive}, the
property is that the parameters of the trained model are biased in a
particular way such that the data contributor can detect her data was
used to train the model.  We will describe more details about the data
manipulation algorithms and their desired properties for the trained
model in \secref{sec:poisoning:prelim} and \secref{sec:tracing:prelim}
when applying our technique to mitigate data poisoning attacks and
counter data tracing, respectively. The data contributor publishes her
manipulated dataset, e.g., posts it on the public Internet such as
social media websites.  Note that the data manipulation algorithm
often preserves utility of the data, e.g., by adding
human-imperceptible perturbation to the data. Moreover, we assume the
\emph{data manipulation algorithm does not modify labels of the data}. This
is because the labels are often assigned by a data curator, which we
discuss next.

\myparagraph{Data curator} A data curator collects data from various
sources, e.g., the public Internet. We call the collected dataset
\emph{manipulated training dataset} and denote it as
$\Dataset{\trainTag}$. In particular, we assume the data contributor's
manipulated dataset is collected by the data curator as a part of
$\Dataset{\trainTag}$.  Moreover, the data curator pre-processes
$\Dataset{\trainTag}$ as a \emph{purified training dataset}
$\Dataset{\purifiedTag}$ and manually labels it if needed. The data
curator then sells $\Dataset{\purifiedTag}$ to model
trainers. Assuming such a data curator is realistic as there are many
of them in the real world, e.g., Argilla (\url{https://argilla.io}),
Cleanlab (\url{https://cleanlab.ai}), and Lightly
(\url{https://www.lightly.ai}).

We assume the data curator has a feature extractor $\Network$, which
outputs a feature vector for an input. For instance, the feature
extractor could be pre-trained on ImageNet~\cite{deng2009:imagenet} or a publicly available one
pre-trained using self-supervised learning; e.g., OpenAI makes its
CLIP feature extractor~\cite{radford2021:learning} pre-trained on 400 million image-text pairs
publicly available (\url{https://openai.com/research/clip}). Moreover,
we assume the data curator has access to a \emph{reference dataset}
$\Dataset{\referenceTag}$.  The data curator could obtain the
reference dataset $\Dataset{\referenceTag}$ from a trustworthy source;
e.g., its employees could generate $\Dataset{\referenceTag}$
themselves. Alternatively, the data curator can automatically detect a
subset of his $\Dataset{\trainTag}$ as $\Dataset{\referenceTag}$ using
an existing detector.  $\Dataset{\referenceTag}$ does not need to be
completely clean due to the imperfection of the detector.  However, we
assume that the fraction of manipulated inputs in
$\Dataset{\referenceTag}$ is much smaller than in
$\Dataset{\trainTag}$. Thus, a model trained based on
$\Dataset{\referenceTag}$ alone is less likely to have the property.
The data curator uses $\Dataset{\referenceTag}$ as a reference to
purify the remaining untrusted dataset
$\Dataset{\trainTag}\setminus\Dataset{\referenceTag}$.

\myparagraph{Model trainer} A model trainer aims to train a model,
e.g., to deploy it as a cloud service or end-user application. Data
collection, labeling, and purification often require a large amount of
data, computation, and human resources, which may not be available to
a model trainer. Thus, we assume the model trainer obtains the
purified training data $\Dataset{\purifiedTag}$ from a data curator
and trains a model based on it, e.g., using a standard supervised
learning algorithm. In this work, we assume the model is a classifier,
which outputs a discrete label in the label set $\ClassesSpace = \{1,
2, \dots, \NumClasses\}$ for an input.

Though we distinguish between a data curator and a model trainer to
make our system setup more general, one party could play both roles.

\subsection{Design Goals} \label{sec:goal}

We aim to design a \emph{data purification framework} for the data
curator to purify his $\Dataset{\trainTag}$ as 
 $\Dataset{\purifiedTag}$. In particular, we aim to
achieve the following three design goals:

\begin{itemize}[nosep,leftmargin=1em,labelwidth=*,align=left]
    \item {\bf Property goal.} Recall that a data contributor
      manipulates her data such that a model trained on her
      manipulated data has some hidden property. Specifically, if the
      data curator shares his manipulated training dataset
      $\Dataset{\trainTag}$ instead of the purified one
      $\Dataset{\purifiedTag}$ with a model trainer, the trained model
      would have the property that the data contributor
      desires. Therefore, the \emph{property goal} means that a model
      trained on the purified training dataset $\Dataset{\purifiedTag}$ does
      \emph{not} have the data-contributor-desired property. For
      instance, in the context of mitigating data poisoning attacks
      performed by a malicious data contributor, the model trained on
      the purified dataset should not misclassify the
      data-contributor-chosen target input; and in the context of
      countering data tracing, the data contributor cannot detect the
      unauthorized use of her data in training the model.

    \item {\bf Utility goal.} The purified training dataset is used to train
      models by model trainers.  A model trainer desires a more
      accurate model. Therefore, the \emph{utility goal} means that
      the purified training dataset has good utility, so that a model trained
      on it is as accurate as possible.

    \item {\bf Generality goal.} A data contributor may use different
      data manipulation algorithms to manipulate her data to achieve
      different properties for the trained model; e.g., data poisoning
      and data tracing represent two popular examples of data
      manipulation algorithms and properties. Moreover, the data
      manipulation algorithms and properties may be unknown to the
      data curator and model trainer. Therefore, the \emph{generality
      goal} means that the data purification framework should be
      agnostic to the data manipulation algorithm and
      data-contributor-desired property.
\end{itemize}

In this work, we aim to design a data purification framework that
achieves these three goals. We only focus on image purification and
leave purification of other types of data as future work.


\section{The \sysName Framework}  \label{sec:purification}

To achieve the property goal, our data purification framework \sysName
aims to perturb the unpurified data to overwhelm the perturbations
added by the data manipulation algorithm of the data contributor, so a
model trained on the purified data does not have the
data-contributor-desired property. A key challenge is how to perturb
the unpurified data to achieve both the property and utility
goals. For instance, one naive method is to add large perturbations to
the unpurified data such that they become random noise.  This method
can achieve the property goal, but it substantially sacrifices the
utility of the purified data. To address this challenge, our key
observation is that the reference dataset achieves the property goal;
i.e., a model trained on the reference dataset is unlikely to have the
data-contributor-desired property, and so a dataset that is
distributed similarly to the reference dataset is likely to achieve
the property goal. Based on this observation, \sysName adds a bounded
perturbation (achieving the utility goal) to the unpurified data such
that the purified data has a feature distribution similar to the
reference data (achieving the property goal), where the features of an
input are extracted by the data curator's feature extractor
$\Network$. We consider feature distribution instead of the raw input
distribution because features are high-level summaries of an input and
are more likely to be relevant to the data-contributor-desired
property of a model.

Formally, we use the well-known Wasserstein distance to quantify the
similarity between two distributions and formulate finding such
perturbations to the unpurified data as an optimization
problem. However, it is challenging to solve the optimization problem
due to the complexity of the Wasserstein distance. To address this
challenge, we propose a two-step method to approximately solve the
optimization problem, which iteratively alternates between optimizing
the perturbations added to the unpurified data and approximating the
Wasserstein distance between the feature distributions of the current
purified data and reference data. \sysName achieves the generality
goal by not depending on any specific data manipulation algorithm and
data-contributor-desired property when formulating the optimization
problem and solving it.

\begin{algorithm*}[!t]
\caption{Our Data Purification Framework \sysName}  \label{ag:purification}
\begin{algorithmic}[1]
\REQUIRE A feature extractor $\Network$, a reference dataset
$\Dataset{\referenceTag}$, an untrusted dataset
$\Dataset{\untrustedTag}$, learning rate $\LearningRate{\Critic}$ for
$\Critic$, learning rate $\LearningRate{\PurifiedPerturbation}$ for
perturbations, purification ratio $\SelectionRatio$, amplification
scale $\Amplification$, gradient penalty
$\GradientPenaltyCoefficient$, and batch size $\BatchSize$.

\STATE $\NumSamples \gets \setSize{\Dataset{\untrustedTag}}$;
\STATE  $\PurifiedPerturbation{\SampleIdx} \leftarrow 0^{\SampleDimension}$, $\forall \SampleIdx=1,2,\cdots, \NumSamples$; \texttt{//Initialize the perturbations}
\STATE  $\Critic \leftarrow$ a random neural network; \texttt{//Initialize neural network $\Critic$}

\WHILE{not meet the stopping criteria}

\STATE \texttt{//Step I: Solve the inner max optimization problem to update $\Critic$.}
\WHILE{not converge}
\STATE \texttt{//Use mini-batches to update $\Critic$.}
\STATE Sample a mini-batch $\{\SampleIdx{\IdxInBatch}\}_{\IdxInBatch=1}^{\BatchSize}$ of $\BatchSize$ indices from $\{1, 2, \dots, \NumSamples\}$;
\STATE Sample a mini-batch  $\{\SampleIdxAlt{\IdxInBatch}\}_{\IdxInBatch=1}^{\BatchSize}$ of $\BatchSize$ indices from $\{1,2,\dots, \setSize{\Dataset{\referenceTag}}\}$;
\STATE $\LossFunction(\Critic) \gets \AveragedFunction(\Critic, \{\Image{}{\SampleIdx{\IdxInBatch}} + \PurifiedPerturbation{\SampleIdx{\IdxInBatch}}\}_{\IdxInBatch=1}^{\BatchSize}) - \AveragedFunction(\Critic, \{\Image{\referenceTag}{\SampleIdxAlt{\IdxInBatch}}\}_{\IdxInBatch=1}^{\BatchSize})$;
\STATE $\Regularization(\Critic) \gets 0$;\label{ag:purification:gradient_penalty:start}
\FOR{$\IdxInBatch=1,2,\cdots,\BatchSize$}  
\STATE Sample a number $\RandomNumber$ from the interval [0,1] uniformly at random;
\STATE $\FeatureVector{}  \gets \RandomNumber \Network(\Image{}{\SampleIdx{\IdxInBatch}} + \PurifiedPerturbation{\SampleIdx{\IdxInBatch}} )$ + $(1-\RandomNumber) \Network(\Image{\referenceTag}{\SampleIdxAlt{\IdxInBatch}})$;
\STATE $\Regularization(\Critic) \gets \Regularization(\Critic) + \frac{1}{\BatchSize} (\euclideanNorm{ \Gradient{\FeatureVector{}} \Critic(\FeatureVector{})} -1)^2$;
\ENDFOR \label{ag:purification:gradient_penalty:end}
\STATE $\Critic \gets \AdamOptimizer(\LossFunction(\Critic)+\GradientPenaltyCoefficient\cdot \Regularization(\Critic), \Critic, \LearningRate{\Critic})$; \texttt{//Update $\Critic$ using the Adam optimizer with learning rate $\LearningRate{\Critic}$}
\ENDWHILE
\STATE \texttt{//Step II: Solve the outer min optimization problem to update perturbations.}
 \STATE Select indices $\DataBatch{\SelectionRatio}{\perturbedTag} \subseteq \{1, \ldots, \NumSamples\}$, $\setSize{\DataBatch{\SelectionRatio}{\perturbedTag}} = \lfloor \SelectionRatio\NumSamples\rfloor$, such that $\forall \SampleIdx \in \DataBatch{\SelectionRatio}{\perturbedTag}, \forall \SampleIdxAlt \not\in \DataBatch{\SelectionRatio}{\perturbedTag} \Rightarrow \Critic(\Image{}{\SampleIdx} + \PurifiedPerturbation{\SampleIdx}) \ge \Critic(\Image{}{\SampleIdxAlt} + \PurifiedPerturbation{\SampleIdxAlt})$;
 \STATE \texttt{//Optimize the perturbation for each selected input.}
 \FOR{$\SampleIdx \in  \DataBatch{\SelectionRatio}{\perturbedTag}$ }
\STATE Compute $\LossFunction(\PurifiedPerturbation{\SampleIdx}) = \Critic(\Network(\Image{}{\SampleIdx} +\PurifiedPerturbation{\SampleIdx})) + \PurificationHyperparameter \euclideanNorm{\PurifiedPerturbation{\SampleIdx}}$;
  \STATE Update perturbation $\PurifiedPerturbation{\SampleIdx}$ using gradient descent:  $\PurifiedPerturbation{\SampleIdx} \gets \PurifiedPerturbation{\SampleIdx} - \LearningRate{\PurifiedPerturbation} \cdot \Gradient{\PurifiedPerturbation{\SampleIdx}}\LossFunction(\PurifiedPerturbation{\SampleIdx})$; \label{ag:purification:perturbation_update}  
  \STATE Clip $\PurifiedPerturbation{\SampleIdx}$ such that $\Image{}{\SampleIdx}+\PurifiedPerturbation{\SampleIdx}$ is within the domain  $\ImageRange^{\SampleDimension}$ if needed;  \label{ag:purification:projection}
  \ENDFOR
\ENDWHILE

\FOR{$\SampleIdx=1,2,\cdots,\NumSamples$}
\STATE $\PurifiedPerturbation{\SampleIdx} \leftarrow  \Amplification\cdot \PurifiedPerturbation{\SampleIdx}$;  \label{ag:purification:perturbation_amplification} \texttt{//Amplify a perturbation.}
 \STATE Clip $\PurifiedPerturbation{\SampleIdx}$ such that $\Image{}{\SampleIdx}+\PurifiedPerturbation{\SampleIdx}$ is within the domain  $\ImageRange^{\SampleDimension}$ if needed;  \label{ag:purification:projection} \texttt{//Project a purified input into the legitimate domain.}
\ENDFOR

\RETURN $\Dataset{\perturbedTag} \gets \{\Image{}{1}+\PurifiedPerturbation{1}, \Image{}{2}+\PurifiedPerturbation{2}, \cdots, \Image{}{\NumSamples}+\PurifiedPerturbation{\NumSamples}\}$
\end{algorithmic}
\end{algorithm*}


\subsection{Formulating an Optimization Problem}

\myparagraph{Formulating an optimization problem} Given a manipulated
training dataset \Dataset{\trainTag}, a reference dataset
$\Dataset{\referenceTag}\subset \Dataset{\trainTag}$, and a feature
extractor $\Network$, we aim to purify the remaining \emph{untrusted
dataset} $\Dataset{\untrustedTag}= \Dataset{\trainTag} \setminus
\Dataset{\referenceTag} = \{\Image{}{1}, \Image{}{2}, \cdots,
\Image{}{\NumSamples}\}$. Our goal is to add a perturbation
$\PurifiedPerturbation{\SampleIdx}$ to each \Image{}{\SampleIdx} such
that the purified untrusted dataset
$\Dataset{\perturbedTag}=\{\Image{}{1}+\PurifiedPerturbation{1},
\Image{}{2}+\PurifiedPerturbation{2}, \cdots,
\Image{}{\NumSamples}+\PurifiedPerturbation{\NumSamples}\}$ achieves
the property and utility goals.  The purified training dataset
$\Dataset{\purifiedTag}$ then consists of both
$\Dataset{\referenceTag}$ and $\Dataset{\perturbedTag}$.  We formulate
finding the perturbations
$\{\PurifiedPerturbation{\SampleIdx}\}_{\SampleIdx=1}^{\NumSamples}$
as the following optimization problem:
\begin{align} \label{eq:min_distance}
    \min_{\{\PurifiedPerturbation{\SampleIdx}\}_{\SampleIdx=1}^{\NumSamples}} & \quad \mathrlap{\DistributionDistance(\Distribution{\perturbedTag}, \Distribution{\referenceTag}) + \PurificationHyperparameter \sum_{\SampleIdx=1}^{\NumSamples} \euclideanNorm{\PurifiedPerturbation{\SampleIdx}}^2} \\ 
    \text{s.t.}\ & \quad \Image{}{\SampleIdx} + \PurifiedPerturbation{\SampleIdx} \in \ImageRange^{\SampleDimension} & \forall{\SampleIdx}=1,2,\cdots,\NumSamples
\end{align}
where $\Distribution{\perturbedTag}$ and
$\Distribution{\referenceTag}$ are respectively the distributions of
the purified data and reference data in the feature space defined by
the feature extractor $\Network$; $\DistributionDistance$ is a
distance metric between two distributions; $\SampleDimension$ is the
number of dimensions of an input/perturbation; and
$\ImageRange^{\SampleDimension}$ is the domain of an input, e.g., the
image domain with pixel values normalized in the range
$\ImageRange$. Both distributions $\Distribution{\perturbedTag}$ and
$\Distribution{\referenceTag}$ are defined in the feature space
$\MetricSet$. Specifically, if we sample an input $\Image{}$ from the
purified dataset (or the reference dataset) uniformly at random, its
feature vector $\Network(\Image{})$ has a probability
$\Distribution{\perturbedTag}(\Network(\Image{}))$ (or
$\Distribution{\referenceTag}(\Network(\Image{}))$) under the
distribution $\Distribution{\perturbedTag}$ (or
$\Distribution{\referenceTag}$).

The first term aims to achieve the property goal, while the second
term aims to achieve the utility goal. Specifically, the first term
aims to perturb the untrusted data such that the purified data has a
similar feature distribution with the reference data. Since a model
trained on the reference data is likely to achieve the property goal,
the purified data, whose features follow a similar distribution with
the reference data, is also likely to achieve the property goal. The
second term aims to minimize the perturbations added to the untrusted
data to preserve its utility. Our optimization problem essentially
achieves a trade-off between the property and utility goals, which can
be tuned by the hyperparameter $\PurificationHyperparameter$.

\myparagraph{Using Wasserstein distance as
  $\DistributionDistance(\Distribution{\perturbedTag},
  \Distribution{\referenceTag})$} The distance
$\DistributionDistance(\Distribution{\perturbedTag},
\Distribution{\referenceTag})$ is a key component of \sysName.
Wasserstein distance offers an interpretable and smooth representation
of the distance between two distributions~\cite{gulrajani2017:improved}. In particular,
Wasserstein distance between two distributions can be interpreted as
the minimum cost of moving a pile of earth in the shape of one
distribution to the other. Therefore, we choose Wasserstein distance
as $\DistributionDistance(\Distribution{\perturbedTag},
\Distribution{\referenceTag})$.  Formally, the Wasserstein distance
$\DistributionDistance(\Distribution{\perturbedTag},
\Distribution{\referenceTag})$ is defined as follows:
\begin{align}   \label{eq:wasserstein_distance}
    \WassersteinDistance(\Distribution{\perturbedTag}, \Distribution{\referenceTag})   \triangleq  \inf_{\JointDistribution \in \JointDistributionSet(\Distribution{\perturbedTag}, \Distribution{\referenceTag})} \expv[(\FeatureVector{}{\perturbedTag},\FeatureVector{}{\referenceTag})\sim\JointDistribution] {\euclideanNorm{\FeatureVector{}{\perturbedTag} - \FeatureVector{}{\referenceTag}}},
\end{align}
where $\JointDistributionSet(\Distribution{\perturbedTag},
\Distribution{\referenceTag})$ is the set of all joint distributions
for a pair of random variables $(\FeatureVector{}{\perturbedTag},
\FeatureVector{}{\referenceTag})$ in the feature space, whose
marginals are $\Distribution{\perturbedTag}$ and
$\Distribution{\referenceTag}$, respectively; $\expv{}$ means
expectation; and
$(\FeatureVector{}{\perturbedTag},\FeatureVector{}{\referenceTag})\sim\JointDistribution$
means that
$(\FeatureVector{}{\perturbedTag},\FeatureVector{}{\referenceTag})$
follows the joint distribution $\JointDistribution$.

\myparagraph{Approximating the Wasserstein distance
  $\WassersteinDistance(\Distribution{\perturbedTag},
  \Distribution{\referenceTag})$} It is a well-known challenge to
compute the Wasserstein distance
$\WassersteinDistance(\Distribution{\perturbedTag},
\Distribution{\referenceTag})$ exactly. In particular, in our problem,
the two distributions do not have analytical forms, making it even
more challenging. To address the challenge, we approximate the
Wasserstein distance in a computationally efficient way.  First,
according to the Kantorovich-Rubinstein duality~\cite{villani2009:optimal}, the
Wasserstein distance can be represented as follows:
\begin{align}
  & \WassersteinDistance(\Distribution{\perturbedTag}, \Distribution{\referenceTag}) \nonumber \\ & \quad \quad = \sup_{{\Critic}\in \CriticSpace} \expv[\FeatureVector{}{\perturbedTag} \sim \Distribution{\perturbedTag}] { \Critic(\FeatureVector{}{\perturbedTag}) } -  \expv[\FeatureVector{}{\referenceTag} \sim \Distribution{\referenceTag}]{ \Critic(\FeatureVector{}{\referenceTag}) },
  \label{eq:min_distance_app}
\end{align}
where $\CriticSpace$ is the set of all $1$-Lipschitz functions in the
feature space.  A function $\Critic: \MetricSet \rightarrow \reals$ is
a $1$-Lipschitz function~\cite{cobzacs2019:lipschitz} if it satisfies the following: 
\begin{align*}
    \euclideanNorm{\Critic(\genericVec) - \Critic(\genericVecAlt)} \leq \euclideanNorm{\genericVec - \genericVecAlt} 
\end{align*}
for all $\genericVec, \genericVecAlt \in \MetricSet$.
Note that $\CriticSpace$ includes all possible $1$-Lipschitz
functions, which makes it still hard to compute
$\WassersteinDistance(\Distribution{\perturbedTag},
\Distribution{\referenceTag})$ using
\eqnref{eq:min_distance_app}. To address the challenge, we
parameterize a $1$-Lipschitz function as a neural network (we use a
two-layer neural network in our experiments) and we approximate
$\CriticSpace$ as $\CriticSpace{\CriticParams}$ that consists of all
$1$-Lipschitz two-layer neural networks. We use a simple two-layer
neural network because the function is defined in the feature space
and to make our method more efficient. Then, we approximate
$\WassersteinDistance(\Distribution{\perturbedTag},
\Distribution{\referenceTag})$ as follows using
$\CriticSpace{\CriticParams}$:

\begin{align*}
     \WassersteinDistance(\Distribution{\perturbedTag}, \Distribution{\referenceTag})\approx\max_{\Critic\in \CriticSpace{\CriticParams}} \expv[\FeatureVector{}{\perturbedTag} \sim \Distribution{\perturbedTag}] {\Critic(\FeatureVector{}{\perturbedTag}) } -  \expv[\FeatureVector{}{\referenceTag} \sim \Distribution{\referenceTag}] {\Critic(\FeatureVector{}{\referenceTag})}.
\end{align*}
The expectation $\expv[\FeatureVector{}{\perturbedTag} \sim
  \Distribution{\perturbedTag}]
{\Critic(\FeatureVector{}{\perturbedTag})}$ can be further
approximated by the average function value of $\Critic$ for the
perturbed data, i.e.,
$\frac{1}{\setSize{\Dataset{\perturbedTag}}}\sum_{\Image\in
  \Dataset{\perturbedTag}}\Critic(\Network(\Image))$, which we denote
as $\AveragedFunction(\Critic, \Dataset{\perturbedTag})$.
Similarly, $\expv[\FeatureVector{}{\referenceTag} \sim
  \Distribution{\referenceTag}]
{\Critic(\FeatureVector{}{\referenceTag})}$ can be approximated as
$\AveragedFunction(\Critic, \Dataset{\referenceTag})$. Thus, we have
$\WassersteinDistance(\Distribution{\perturbedTag},
\Distribution{\referenceTag}) \approx
\max_{\Critic\in \CriticSpace{\CriticParams}}
\AveragedFunction(\Critic, \Dataset{\perturbedTag}) -
\AveragedFunction(\Critic, \Dataset{\referenceTag})$.

\myparagraph{Refining the optimization problem} After approximating
$\WassersteinDistance(\Distribution{\perturbedTag},
\Distribution{\referenceTag})$, we can reformulate
\eqnref{eq:min_distance} as the following optimization problem:
\begin{align} 
  \min_{\{\PurifiedPerturbation{\SampleIdx}\}_{\SampleIdx=1}^{\NumSamples}} & \max_{\Critic\in \CriticSpace{\CriticParams}}  \AveragedFunction(\Critic, \Dataset{\perturbedTag}) - \AveragedFunction(\Critic, \Dataset{\referenceTag}) + \PurificationHyperparameter\sum_{\SampleIdx=1}^{\NumSamples} \euclideanNorm{\PurifiedPerturbation{\SampleIdx}}^2
  \label{eq:minimax:goal} \\ 
  \text{s.t.}\ & \Image{}{\SampleIdx} + \PurifiedPerturbation{\SampleIdx} \in \ImageRange^{\SampleDimension} \quad\quad \forall \SampleIdx=1,2,\cdots,\NumSamples. 
  \label{eq:minimax:constraints}
\end{align}
Note that the reformulated optimization problem is a min-max one with
an inner maximization optimization and an outer min optimization.

\subsection{Solving the Optimization Problem}

It is challenging to exactly solve the min-max optimization problem in
\eqnsref{eq:minimax:goal}{eq:minimax:constraints} due to the
complexity of the neural network $\Critic$. To address this challenge,
we propose an iterative two-step solution to approximately solve it,
which is shown in~\algref{ag:purification}. Specifically, we first
initialize each perturbation $\PurifiedPerturbation{\SampleIdx}$ as
$0^{\SampleDimension}$ and $\Critic$ as a random neural network. Then,
we iteratively alternate between Step I and Step II, where Step I aims
to solve the inner max problem to update $\Critic$ and Step II aims to
solve the outer min problem to update the perturbations.

\myparagraph{Step I: Solve the inner max problem} Given the current
perturbations, Step I solves the inner max problem to update
$\Critic$. Formally, the inner max problem is as follows:
$\max_{\Critic\in \CriticSpace{\CriticParams}}
\LossFunction(\Critic)\triangleq \AveragedFunction(\Critic,
\Dataset{\perturbedTag}) - \AveragedFunction(\Critic,
\Dataset{\referenceTag})$. We use a gradient ascent based method to
iteratively solve the max problem. Specifically, in each iteration, we
sample a mini-batch of $\Dataset{\perturbedTag}$ and a mini-batch of
$\Dataset{\referenceTag}$ to estimate $\LossFunction(\Critic)$; and
then we use the Adam optimizer~\cite{kingma2014:adam} to update
$\Critic$. However, a key challenge is that $\Critic$ should be a
$1$-Lipschitz function, i.e.,
$\Critic\in \CriticSpace{\CriticParams}$. The updated $\Critic$ in the
above iterative process does not necessarily guarantee that $\Critic$
is a $1$-Lipschitz function. To address this challenge, we add a
penalty term to the loss function $\LossFunction(\Critic)$, which is
inspired by prior work on approximating Wasserstein
distance~\cite{gulrajani2017:improved}. The key idea of the penalty
term is to reduce the magnitude of the gradient of the function
$\Critic$ with respect to its input, so the updated $\Critic$ is more
likely to be a $1$-Lipschitz
function. \linesref{ag:purification:gradient_penalty:start}{ag:purification:gradient_penalty:end}
in~\algref{ag:purification} correspond to calculating the penalty
term, where $\GradientPenaltyCoefficient$ is a penalty coefficient
used to balance $\LossFunction(\Critic)$ and the penalty term.

We repeat the iterative process until the loss function
$\LossFunction(\Critic)$ converges. Specifically, we save the value of
$\LossFunction(\Critic)$ calculated using the entire
$\Dataset{\perturbedTag}$ and $\Dataset{\referenceTag}$ as a
checkpoint in every $\CheckingPoint$ steps. If the relative difference
between $\LossFunction(\Critic)$ in the current checkpoint and the
average $\LossFunction(\Critic)$ in the latest $5$ checkpoints is less
than $1\%$, we stop the iterative process.

\myparagraph{Step II: Solve the outer min problem} Given the current
$\Critic$, Step II solves the outer min problem to update the
perturbations. Instead of updating each perturbation
$\PurifiedPerturbation{\SampleIdx}$, we only update those whose
corresponding purified inputs are far away from the centroid of the
distribution of the reference data, i.e., have high function values
under the current $\Critic$. We do this to enhance the utility of the
purified data. Specifically, we select the $\SelectionRatio$ fraction
of inputs in the purified dataset that have the largest function
values under $\Critic$.
For each selected purified input
$\Image{}{\SampleIdx} + \PurifiedPerturbation{\SampleIdx}$, we update
its $\PurifiedPerturbation{\SampleIdx}$. Specifically, for a
perturbation $\PurifiedPerturbation{\SampleIdx}$, the inner min
problem becomes: $ \min_{\PurifiedPerturbation{\SampleIdx}}
\Critic(\Network(\Image{}{\SampleIdx} +
\PurifiedPerturbation{\SampleIdx})).$ We use gradient
descent to update $\PurifiedPerturbation{\SampleIdx}$ for one
iteration (\Lineref{ag:purification:perturbation_update}
in~\algref{ag:purification}).

\myparagraph{Stopping criteria, perturbation amplification, and
  projection} We alternate between Step I and Step II until a stopping
criteria is satisfied. Specifically, we stop the iterative process
when the number of rounds reaches a predefined maximum number of
rounds (we use 30 in experiments) or the objective function values of
the min-max problem in the latest 5 rounds are all larger than the
minimal objective function value seen so far.  After stopping the
iterative process, we amplify each perturbation
(\Lineref{ag:purification:perturbation_amplification}) to better
achieve the property goal~\cite{chandrasekaran2021face}. Finally, we clip the perturbation such that
each purified input is within the legitimate domain
$\ImageRange^{\SampleDimension}$
(\Lineref{ag:purification:projection}).


\section{As a Defense Against Data Poisoning}  \label{sec:poisoning}

\subsection{Clean-Label Data Poisoning} \label{sec:poisoning:prelim}

Generally, clean-label data
poisoning~\cite{huang2020:metapoison,geiping2020:witches,souri2022:sleeper} aims to
perturb a subset of training inputs without changing their semantics
such that a model trained on a correctly labeled dataset containing
the perturbed inputs would misclassify an attacker-chosen input
(called the \emph{target input}) or any input
with a attacker-chosen patch (called \emph{trigger}) into an attacker-chosen, incorrect
label (called the \emph{target label}).  A data manipulation algorithm
in this context takes a labeled training dataset, a target input
\TargetImage (or a trigger \Trigger), and a target label \TargetLabel
as input and outputs a poisoned training dataset. The data-contributor-desired
property is that a model trained on the poisoned training dataset
should predict \TargetLabel for \TargetImage (\emph{targeted attack}) or
predict any input with the trigger \Trigger for \TargetLabel (\emph{backdoor attack}).
We consider both \emph{targeted attack} and \emph{backdoor attack} in this work.

In particular, a bounded perturbation is added to a subset of training
inputs to preserve their semantics. The key challenge of data
poisoning is what perturbations should be added to the training
inputs.  This challenge is often addressed by formulating finding the
perturbations as a bi-level optimization problem~\cite{biggio2012:poisoning,munoz2017:towards}, where the
inner optimization problem is to minimize the training loss (i.e.,
inner objective function) of a model on the poisoned training dataset,
while the outer optimization problem is to minimize the loss (i.e.,
outer objective function) of the model for the
(\TargetImage, \TargetLabel) pair in \emph{targeted attack} 
(or (\Trigger, \TargetLabel) pair in \emph{backdoor attack}). 
Different data poisoning attacks use different heuristics to approximately solve the
bi-level optimization problem. For instance, \textit{Gradient
  Matching}~\cite{geiping2020:witches,souri2022:sleeper}, the state-of-the-art clean-label data poisoning attack,
uses a gradient based method to iteratively solve the bi-level
optimization problem; and in each iteration, it aims to update the
perturbations to maximize the alignment between the gradient of the
inner objective function and that of the outer objective
function. \textit{Gradient Matching} adds {perturbations} within a ball
of radius $\PoisonHyperparameter$ in the infinity norm to a subset of
training inputs. In our experiments, we use
\textit{Gradient Matching}~\cite{geiping2020:witches,souri2022:sleeper} to manipulate data.

\subsection{Experimental Setup}  \label{sec:poisoning:setup}

\myparagraph{Datasets} 
We used two visual benchmarks: CIFAR-10~\cite{krizhevsky2009:learning} and TinyImageNet~\cite{Lle2015:TinyIV}. 
\begin{itemize}[nosep,leftmargin=1em,labelwidth=*,align=left]
\item \textbf{CIFAR-10.} CIFAR-10 is a dataset containing $60{,}000$
  images, each of dimension $32\times32\times3$, partitioned into
  $\NumClasses=10$ classes in which $50{,}000$ images are training
  samples and $10{,}000$ are test samples.
\item \textbf{TinyImageNet.} TinyImageNet is a dataset containing
  images of dimension $64\times64\times3$, partitioned into
  $\NumClasses=200$ classes. In each class, there are $500$ training
  images and $50$ validation images. We sampled $50$ classes (see
  \appref{sec:app:sample}) and used the data from these $50$ classes
  in our experiments. As such, there are $25{,}000$ training images of
  $50$ classes and $2{,}500$ testing images.
\end{itemize}

\myparagraph{Poisoning setting} We setup the targeted attack setting
represented by $\langle\Dataset{\trainTag},
\TargetImage, \TargetLabel\rangle$ in one experiment as
follows.  We first sampled one label from $10$ classes in CIFAR-10, or
from $50$ classes in TinyImageNet, uniformly at random as the target
label $\TargetLabel$.  Then, we sampled one testing image whose
label is not $\TargetLabel$ uniformly at random as the target
input $\TargetImage$.  Next, we used Gradient Matching to
generate a poisoned training dataset $\Dataset{\trainTag}$.
Specifically, following previous work~\cite{geiping2020:witches}, we
uniformly sampled $1\%$ of training images from the target class
(i.e., $500$ samples in CIFAR-10 and $250$ samples in TinyImageNet) to
be poisoned, where we set $\PoisonHyperparameter=16$ (out of the 255
maximum pixel value). Following the implementation by the previous
work\footnote{\url{https://github.com/JonasGeiping/poisoning-gradient-matching}},
we constructed $20$ sets of $\langle\Dataset{\trainTag},
\TargetImage, \TargetLabel\rangle$ for CIFAR-10 and
another $20$ sets of $\langle\Dataset{\trainTag},
\TargetImage, \TargetLabel\rangle$ for TinyImageNet.

We setup the backdoor setting represented by
$\langle\Dataset{\trainTag}, \Trigger, \TargetLabel\rangle$ in one
experiment as follows.  We first sampled one label from $10$ classes in
CIFAR-10, or from $50$ classes in TinyImageNet, uniformly at random as
the target label $\TargetLabel$. We used the same patch from the
previous work~\cite{souri2022:sleeper} as \Trigger. Next, we used
Gradient Matching to generate a poisoned training dataset
$\Dataset{\trainTag}$.  Specifically, following previous
work~\cite{souri2022:sleeper}, we uniformly sampled $1\%$ of training
images from the target class (i.e., $500$ samples in CIFAR-10 and
$250$ samples in TinyImageNet) to be poisoned, where we set
$\PoisonHyperparameter=16$ (out of the 255 maximum pixel
value). Following the implementation by the previous
work\footnote{\url{https://github.com/hsouri/Sleeper-Agent}}, we
constructed $20$ sets of $\langle\Dataset{\trainTag}, \Trigger,
\TargetLabel\rangle$ for CIFAR-10 and another $20$ sets of
$\langle\Dataset{\trainTag}, \Trigger, \TargetLabel\rangle$ for
TinyImageNet.

\myparagraph{Selecting reference data}  
 \sysName requires a detector to detect a reference dataset
$\Dataset{\referenceTag}$ from a poisoned training dataset
$\Dataset{\trainTag}$.  We consider both \emph{simulated detectors}
and a \emph{real-world detector} for a comprehensive evaluation of
\sysName in different scenarios. A simulated detector enables us to
explicitly control the performance of the detector and thus the
quality of the reference dataset, which makes it possible to
understand the performance of \sysName in different scenarios. A
real-world detector is a state-of-the-art detector that is currently
available to a data curator, which enables us to understand the
performance of \sysName as of now.
\begin{itemize}[nosep,leftmargin=1em,labelwidth=*,align=left]
\item \textbf{Simulated detectors.} We characterize a simulated detector
using its \emph{precision} $\Precision$ and \emph{recall} $\Recall$. Precision is the
fraction of inputs in the detected reference dataset
$\Dataset{\referenceTag}$ that are unmanipulated, while recall is the
fraction of unmanipulated inputs in the poisoned training dataset
$\Dataset{\trainTag}$ that are detected to be in
$\Dataset{\referenceTag}$. Given a precision $\Precision$ and recall
$\Recall$, we uniformly sampled a subset of size
$\floor{\setSize{\Dataset{\trainTag}}\times0.99\times\Recall\times\frac{1-\Precision}{\Precision}}$
from the poisoned/manipulated inputs and a subset of size
$\floor{\setSize{\Dataset{\trainTag}}\times0.99\times\Recall}$ from the
unmanipulated inputs, and merged them to create
$\Dataset{\referenceTag}$. Note that 0.99 appears in these
calculations because the fraction of poisoned/manipulated inputs is
1\%. When the precision of this detector is $1.0$, this simulates the
scenario where reference data is collected from trusted sources. 
Note that 1\% of the training data is poisoned. Therefore, we consider
the precision $\Precision>0.99$ so that the simulated detector is better 
than the random detector that selects the reference data from the poisoned 
training dataset uniformly at random.

\item \textbf{Real-world detectors.} This type of detector represents an
algorithm available as of now that can be used to detect reference
data. In our experiments, we considered the state-of-the-art detection method
known as $\knnParameter$-Nearest Neighbors
($\knnParameter$-NN)~\cite{peri2020:deep}. The previous work~\cite{peri2020:deep} demonstrated
its effectiveness in the scenario where DNN is trained by fine-tuning
the linear classifier with the feature extractor frozen. However, it
is less effective to remove all the data that introduces poison into a
DNN trained from scratch (please see the results in \tblref{table:poisoning:baselines}). 
Therefore, we adapted the $\knnParameter$-NN such that the data selected by it are
mostly unmanipulated, enabling this data to be used as
\Dataset{\referenceTag} in purification.
We denote its adapted version as $\knnParameter$-NN($\NumThreshold$) where 
$\NumThreshold$ is number of detected samples per class. 
The details of $\knnParameter$-NN($\NumThreshold$) are described in \appref{app:adapted_knn}.

As suggested by the previous work~\cite{peri2020:deep}, we set
$\knnParameter$ as the number of training samples per class, i.e.,
$\knnParameter = 5000$ for CIFAR-10 and $\knnParameter = 500$ for
TinyImageNet. We explored different options of $\NumThreshold$ in
$\knnParameter$-NN($\NumThreshold$).

\end{itemize}

\myparagraph{Purification setting} After detecting the reference
dataset $\Dataset{\referenceTag}$, we used \sysName to purify the
remaining untrusted data $\Dataset{\untrustedTag}=\Dataset{\trainTag}
\setminus \Dataset{\referenceTag}$.  \sysName was set to use ResNet-18
pretrained on ImageNet as the default feature extractor $\Network$.
We set $\GradientPenaltyCoefficient = 10$; $\BatchSize = 256$;
$\LearningRate{\Critic} = 0.001$;
$\LearningRate{\PurifiedPerturbation} = 2.0$ for CIFAR-10 and
$\LearningRate{\PurifiedPerturbation} = 10.0$ for TinyImageNet;
$\SelectionRatio = 5\%$ for CIFAR-10 and $\SelectionRatio = 10\%$ for
TinyImageNet; $\Amplification=2.0$; and the interval of checkpoints
$\CheckingPoint = 20$ for CIFAR-10 and $\CheckingPoint = 1$ for
TinyImageNet.  Instead of choosing the regularization parameter
$\PurificationHyperparameter>0$ to constrain the perturbation and
alternating our two steps for many iterations until convergence, we
set it to be $0$ but applied an ``early stopping'' strategy to bound
the perturbation and reduce computation overhead, which is a commonly
used technique in machine learning.  In particular, we set the max
number of iterations to be $30$. 

\myparagraph{Setting for training models} We trained ResNet-18 models
on normalized data with default data augmentation
applied~\cite{geiping2020:witches} using an SGD
optimizer~\cite{amari1993:backpropagation} for $40$ epochs with a
batch size of $128$ and an initial learning rate of $0.1$ decayed by a
factor of $0.1$ when the number of epochs reaches $15$, $25$, or
$35$. When using a purified training dataset to train a model, the
detected reference dataset was used as a part of the purified training
dataset.

\myparagraph{Evaluation metrics} 
We used \emph{accuracy} ($\Accuracy$) and \emph{attack success rate}
($\AttackSuccessRate$) to measure a trained model, where $\Accuracy$
is the fraction of testing samples (excluding $(\TargetImage,
\TargetLabel)$ in a \emph{targeted attack}) that are correctly
classified by a model and $\AttackSuccessRate$ is the fraction of the
target inputs in a \emph{targeted attack}, or inputs patched with the
trigger in a \emph{backdoor attack}, that are classified as the target
labels by a model. A higher $\Accuracy$ and a lower
$\AttackSuccessRate$ indicate a better performance of a data
purification method.

\subsection{Baselines}  \label{sec:poisoning:baseline}

\myparagraph{Reference} 
The first method, denoted as $\ReferenceMethod$, is to simply remove
the remaining untrusted data after detecting the reference data. That is, it
returns $\Dataset{\referenceTag}$ as $\Dataset{\purifiedTag}$ where
$\Dataset{\referenceTag}$ is the set of reference data detected by a
detector from $\Dataset{\trainTag}$. Then, models are trained using
the reference data alone.

\myparagraph{Purification}
We considered two purification methods as baselines: \textit{Random
  Noise}, denoted as $\RandomNoise{\GassianVariant}$, constructs
perturbations sampled from a Gaussian distribution with a standard
deviation of $\GassianVariant$; and \textit{Friendly
  Noise}~\cite{liu2022:friendly}, denoted as
$\FriendlyNoise{\PerturbationMagnitude}$, constructs a perturbation
that is maximized within a ball of radius $\PerturbationMagnitude$ in
the infinity norm but that, when added to the data, does not
substantially change the model's output. We explored $\GassianVariant
\in [16, \infty)$ and $\PerturbationMagnitude \in [16, \infty)$.

\myparagraph{Prevention}  
We considered four prevention methods as baselines:
\FriendlyNoiseB\footnote{\FriendlyNoiseB adds friendly noise into
training data and introduces Bernoulli noise into the training
process. Since it requires changes in the learning algorithm, we
considered it as a prevention method.}~\cite{liu2022:friendly} and
\EPIC~\cite{yang2022:not}, and two recently proposed defenses against
dirty-label backdoor attacks: \ASD~\cite{gao2023:backdoor} and
\CBD~\cite{zhang2023:backdoor}.

The descriptions of baselines and their implementation details 
are introduced in \appref{app:baseline}.

\subsection{Experimental  Results}  \label{sec:poisoning:results}

\myparagraph{Results without purification} 
\tblref{table:poisoning:performance} shows the average $\Accuracy$ and
$\AttackSuccessRate$ of the models trained on the poisoned training
datasets without purification. Our results show that the attacks are very successful as the $\AttackSuccessRate$s are
high. We note that our results match well with previous works~\cite{geiping2020:witches,souri2022:sleeper}.

\begin{table}[t!]
  \centering
  \vspace{0.1in}
  \resizebox{0.476\textwidth}{!}{
    \begin{tabular}{@{}lr@{\hspace{1.5em}}rr@{\hspace{1.5em}}rr@{\hspace{1.5em}}r@{}}
    \toprule
    \multicolumn{1}{c}{\multirow{2}{*}{}} &  \multicolumn{2}{c}{Targeted} & \multicolumn{2}{c}{Backdoor} \\
     & \multicolumn{1}{c}{\Accuracy \%} & \multicolumn{1}{c}{\AttackSuccessRate} & \multicolumn{1}{c}{\Accuracy \%} & \multicolumn{1}{c}{\AttackSuccessRate\%}  \\
    \midrule
    CIFAR-10 &  $91.97(\pm0.25)$ & $17/20$ & $91.93(\pm0.39)$ & $89.47$ \\
    TinyImageNet & $61.32(\pm0.95)$ & $18/20$ & $61.08(\pm0.79)$ & $81.09$\\
    \bottomrule
    \end{tabular}
    }
      \caption{Average \Accuracy and \AttackSuccessRate of the models
        trained on the poisoned datasets. The numbers in the parenthesis
        are standard deviations among the 20 experiment trials.}
      \label{table:poisoning:performance}
  \end{table}

  \begin{table*} [ht!]
    \centering
    \vspace{0.1in}
    {\resizebox{\textwidth}{!}{
        \begin{tabular}{@{}lr@{\hspace{1.5em}}rr@{\hspace{1.5em}}rr@{\hspace{1.5em}}rr@{\hspace{1.5em}}rr@{\hspace{1.5em}}rr@{\hspace{1.5em}}rr@{\hspace{1.5em}}r@{}}
        \toprule
            \multicolumn{1}{c}{\multirow{2}{*}{}} & \multicolumn{1}{c}{} & \multicolumn{2}{c}{$\Precision=0.991$} & \multicolumn{2}{c}{$\Precision=0.993$} & \multicolumn{2}{c}{$\Precision=0.995$}  & \multicolumn{2}{c}{$\Precision=0.998$}  & \multicolumn{2}{c}{$\Precision=1.0$} \\
              & & \multicolumn{1}{c}{\Accuracy \%} & \multicolumn{1}{c}{\AttackSuccessRate} & \multicolumn{1}{c}{\Accuracy \%} & \multicolumn{1}{c}{\AttackSuccessRate} & \multicolumn{1}{c}{\Accuracy \%} & \multicolumn{1}{c}{\AttackSuccessRate}  & \multicolumn{1}{c}{\Accuracy \%} & \multicolumn{1}{c}{\AttackSuccessRate}  & \multicolumn{1}{c}{\Accuracy \%} & \multicolumn{1}{c}{\AttackSuccessRate}  \\
        \midrule
        
        \multirow{4}{*}{$\Recall=0.1$}& \ReferenceMethod &$56.22(\pm3.96)$ & $0/20$ & $56.90(\pm4.52)$ & $1/20$ & $55.28(\pm4.55)$ & $0/20$ & $56.09(\pm3.32)$ & $0/20$ & $56.17(\pm3.10)$ & $1/20$ \\
       & \RandomNoise & $\mathbf{90.54(\pm0.41)}$ & $5/20$ & $\mathbf{90.58(\pm0.42)}$ & $2/20$ & $90.48(\pm0.31)$ & $2/20$ & $\mathbf{90.44(\pm0.36)}$ & $3/20$ & $\mathbf{90.59(\pm0.26)}$ & $4/20$  \\
       & \FriendlyNoise & $\mathbf{90.49(\pm0.39)}$ & $\mathbf{2/20}$ & $\mathbf{90.49(\pm0.25)}$ & $2/20$ & $90.41(\pm0.38)$ & $4/20$ & $\mathbf{90.62(\pm0.42)}$ & $2/20$ & $\mathbf{90.55(\pm0.34)}$ & $2/20$  \\
        & \sysName & $\mathbf{91.47(\pm0.35)}$ & $\mathbf{2/20}$ & $\mathbf{91.47(\pm0.36)}$ & $\mathbf{0/20}$ & $\mathbf{91.63(\pm 0.43)}$ & $\mathbf{1/20}$ & $\mathbf{91.32(\pm0.53)}$ & $\mathbf{0/20}$ & $\mathbf{91.44(\pm0.42)}$ & $\mathbf{0/20}$ \\  \midrule
    
        \multirow{4}{*}{$\Recall=0.2$}& \ReferenceMethod & $72.77(\pm2.32)$ & $0/20$ & $72.47(\pm3.03)$ & $0/20$ & $72.88(\pm2.40)$ & $0/20$ & $72.05(\pm2.81)$ & $1/20$ & $73.59(\pm3.24)$ & $0/20$ \\
        & \RandomNoise & $90.61(\pm0.33)$ & $3/20$ & $90.70(\pm0.32)$ & $4/20$ & ${90.71(\pm0.28)}$ & ${1/20}$ & $90.65(\pm0.30)$ & $4/20$ & $90.85(\pm0.30)$ & $2/20$  \\
        & \FriendlyNoise & $90.70(\pm0.26)$ & $5/20$ & $90.76(\pm0.37)$ & $3/20$ & $90.68(\pm0.24)$ & $4/20$ & $90.79(\pm0.33)$ & $3/20$ & ${90.73(\pm0.32)}$ & ${1/20}$  \\
        & \sysName & $\mathbf{91.93(\pm0.34)}$ & $\mathbf{5/20}$ & $\mathbf{91.85(\pm0.24)}$ & $\mathbf{7/20}$  & $\mathbf{91.86(\pm0.36)}$ & $\mathbf{4/20}$ & $\mathbf{91.81(\pm0.40)}$ & $\mathbf{0/20}$ & $\mathbf{91.93(\pm0.44)}$ & $\mathbf{1/20}$ \\  \midrule
    
        \multirow{4}{*}{$\Recall=0.3$}& \ReferenceMethod & $76.19(\pm2.35)$ & $1/20$ & $80.89(\pm2.13)$ & $1/20$ & $80.70(\pm1.70)$ & $1/20$ & $80.96(\pm1.24)$ & $0/20$ & $71.95(\pm3.80)$ & $0/20$ \\
        & \RandomNoise & $91.01(\pm0.22)$ & $4/20$ & $90.97(\pm0.29)$ & $4/20$ & $\mathbf{90.97(\pm0.30)}$ & $\mathbf{2/20}$ & $90.96(\pm0.28)$ & $2/20$ & $\mathbf{90.97(\pm0.29)}$ & $5/20$  \\
        & \FriendlyNoise & $90.86(\pm0.29)$ & $3/20$  & $90.80(\pm0.28)$ & $3/20$ & $90.90(\pm0.23)$ & $3/20$ & $90.95(\pm0.28)$ & $4/20$ & $90.90(\pm0.31)$ & $4/20$  \\
        & \sysName & $\mathbf{92.02(\pm0.38)}$ & $\mathbf{7/20}$  & $\mathbf{92.09(\pm0.25)}$ & $\mathbf{7/20}$ & $\mathbf{91.73(\pm0.52)}$ & ${4/20}$ & $\mathbf{92.03(\pm0.31)}$ & $\mathbf{1/20}$ & $\mathbf{91.96(\pm0.46)}$  & $\mathbf{0/20}$ \\  
    
        \bottomrule
        \end{tabular}
        }}
    
        \caption{Experimental results of purification against \emph{targeted data
          poisoning} on CIFAR-10 where simulated detectors parameterized by precision
          $\Precision$ and recall $\Recall$ were applied. \GassianVariant
          and \PerturbationMagnitude in baselines were searched such that
          their \Accuracy is as close as possible to \Accuracy of
          \sysName. In each \Accuracy column, we \textbf{bold} the largest
          number and those within 1.0\%, indicating that those methods
          achieved comparable \Accuracy; and in each \AttackSuccessRate
          column, we \textbf{bold} the smallest number(s) among the
          methods whose \Accuracy is bold (i.e., achieving comparable
          \Accuracy). A method with bold both \Accuracy and
          \AttackSuccessRate means that it achieved the smallest
          \AttackSuccessRate when achieving comparable \Accuracy with
          others.  }
        \label{table:poisoning:human:cifar10}
    \end{table*}
    
    \begin{table*} [ht!]
    \centering
    \vspace{0.1in}
    {\resizebox{\textwidth}{!}{
        \begin{tabular}{@{}lr@{\hspace{0.3em}}rr@{\hspace{0.3em}}rr@{\hspace{0.3em}}rr@{\hspace{0.3em}}rr@{\hspace{0.3em}}rr@{\hspace{0.3em}}r@{}}
        \toprule
            \multicolumn{1}{c}{\multirow{2}{*}{}} & \multicolumn{2}{c}{$\NumThreshold=500$} & \multicolumn{2}{c}{$\NumThreshold=1000$} & \multicolumn{2}{c}{$\NumThreshold=2000$}  & \multicolumn{2}{c}{$\NumThreshold=3000$}  & \multicolumn{2}{c}{$\NumThreshold=4000$} \\
               & \multicolumn{1}{c}{\Accuracy \%} & \multicolumn{1}{c}{$\AttackSuccessRate\quad$} & \multicolumn{1}{c}{\Accuracy \%} & \multicolumn{1}{c}{$\AttackSuccessRate\quad$} & \multicolumn{1}{c}{\Accuracy \%} & \multicolumn{1}{c}{$\AttackSuccessRate\quad$}  & \multicolumn{1}{c}{\Accuracy \%} & \multicolumn{1}{c}{$\AttackSuccessRate\quad$}  & \multicolumn{1}{c}{\Accuracy \%} & \multicolumn{1}{c}{$\AttackSuccessRate\quad$}  \\
        \midrule
        \ReferenceMethod & $51.71 (\pm 2.45)$ & $1/20$ & $66.46(\pm1.62)$ & $0/20$ & $77.79(\pm0.60)$ & $0/20$ & $84.43(\pm0.56)$ & $1/20$ & $88.83(\pm 0.44)$ & $3/20$ \\
        \RandomNoise & $\mathbf{86.54(\pm 0.18)}$ & $\mathbf{0/20}$ & $\mathbf{86.66(\pm0.33)}$ & $\mathbf{0/20}$ & $\mathbf{88.57(\pm0.30)}$ & $\mathbf{0/20}$ & $\mathbf{89.11(\pm0.27)}$ & $\mathbf{0/20}$ & $\mathbf{90.35(\pm0.31)}$ & $3/20$ \\
        \FriendlyNoise & $\mathbf{86.99(\pm0.27)}$ & $\mathbf{0/20}$ & $\mathbf{86.52(\pm0.29)}$ & $\mathbf{0/20}$ & $\mathbf{88.60(\pm0.40)}$ & $\mathbf{0/20}$ & $\mathbf{89.63(\pm0.32)}$ & $\mathbf{0/20}$ & $\mathbf{90.87(\pm0.31)}$ & $3/20$ \\
        \sysName & $\mathbf{86.25(\pm0.45)}$ & $\mathbf{0/20}$ & $\mathbf{86.93(\pm0.32)}$ & $\mathbf{0/20}$ & $\mathbf{88.23(\pm0.37)}$ & $\mathbf{0/20}$ & $\mathbf{89.49(\pm0.35)}$ & $\mathbf{0/20}$ & $\mathbf{90.86(\pm0.27)}$ & $3/20$ \\
        \bottomrule
        \end{tabular}
        }}
    
        \caption{Experimental results of purification against \emph{targeted data
          poisoning} on CIFAR-10 where \knnParameter-NN(\NumThreshold) detectors were
          applied. The precision and recall of such detectors are shown in
          \tblref{table:poisoning:knn-kappa:cifar10:parameters} in
          \appref{sec:app:knn}. We bold the numbers following the same
          rule as \tblref{table:poisoning:human:cifar10}.}
        \label{table:poisoning:knn-kappa:cifar10}
    \end{table*}


 \begin{table*} [ht!]
  \centering
  \vspace{0.1in}
  {\resizebox{\textwidth}{!}{
      \begin{tabular}{@{}lr@{\hspace{1.8em}}rr@{\hspace{1.8em}}rr@{\hspace{1.8em}}rr@{\hspace{1.8em}}rr@{\hspace{1.8em}}rr@{\hspace{1.8em}}rr@{\hspace{1.8em}}r@{}}
      \toprule
          \multicolumn{1}{c}{\multirow{2}{*}{}} & \multicolumn{1}{c}{} & \multicolumn{2}{c}{$\Precision=0.991$} & \multicolumn{2}{c}{$\Precision=0.993$} & \multicolumn{2}{c}{$\Precision=0.995$}  & \multicolumn{2}{c}{$\Precision=0.998$}  & \multicolumn{2}{c}{$\Precision=1.0$} \\
            & & \multicolumn{1}{c}{$\Accuracy\%$} & \multicolumn{1}{c}{$\AttackSuccessRate\%$} & \multicolumn{1}{c}{$\Accuracy\%$} & \multicolumn{1}{c}{$\AttackSuccessRate\%$} & \multicolumn{1}{c}{$\Accuracy\%$} & \multicolumn{1}{c}{$\AttackSuccessRate\%$}  & \multicolumn{1}{c}{$\Accuracy\%$} & \multicolumn{1}{c}{$\AttackSuccessRate\%$}  & \multicolumn{1}{c}{\Accuracy \%} & \multicolumn{1}{c}{$\AttackSuccessRate\%$}  \\
      \midrule
      
      \multirow{4}{*}{$\Recall=0.1$}& \ReferenceMethod & $58.55(\pm4.15)$ & $6.24$ & $57.54(\pm4.11)$ & $7.20$ & $57.26(\pm 3.84)$ & $6.13$ & $56.80(\pm3.71)$ & $5.91$ & $56.05(\pm3.89)$ & $6.55$ \\  
      & \RandomNoise & $\mathbf{90.35(\pm0.42)}$ & $35.73$ & $90.51(\pm0.40)$ & $32.41$ & $\mathbf{90.47(\pm0.27)}$ & $35.86$ & $\mathbf{90.50(\pm0.36)}$ & $32.61$ & $\mathbf{90.55(\pm0.25)}$ & $35.76$\\
      & \FriendlyNoise & $89.81(\pm0.39)$ & $31.58$ & $89.81(\pm0.42)$ & $32.64$ & $89.91(\pm0.33)$ & $34.45$ & $89.86(\pm0.43)$ & $29.69$ & $89.88(\pm0.41)$ & $37.15$ \\
      & \sysName & $\mathbf{91.20(\pm0.67)}$ & $\mathbf{7.48}$ & $\mathbf{91.57(\pm0.46)}$ & $\mathbf{10.34}$ & $\mathbf{91.36(\pm 0.39)}$ & $\mathbf{12.02}$ & $\mathbf{91.33(\pm0.54)}$ & $\mathbf{5.03}$ & $\mathbf{91.30(\pm0.37)}$ & $\mathbf{2.76}$ \\  \midrule
  
      \multirow{4}{*}{$\Recall=0.2$} & \ReferenceMethod & $72.84(\pm3.21)$ & $5.45$ & $73.23(\pm3.48)$ & $5.05$  & $71.68(\pm3.17)$ & $4.46$ & $73.55(\pm2.55)$ & $3.09$ & $72.50(\pm2.90)$ & $3.28$ \\  
      & \RandomNoise  & $90.63(\pm0.35)$ & $35.59$ & $90.71(\pm0.31)$ & $36.24$ & $90.66(\pm0.44)$ & $36.81$ & $90.70(\pm0.36)$ & $39.25$ & $90.70(\pm0.28)$ & $35.55$\\
      & \FriendlyNoise & $90.18(\pm0.43)$ & $38.5$ & $90.15(\pm0.45)$ & $35.8$ & $90.17(\pm0.38)$ & $36.33$ & $90.13(\pm0.40)$ & $33.5$ & $90.24(\pm0.36)$ & $34.95$\\
      & \sysName & $\mathbf{91.83(\pm0.39)}$ & $\mathbf{23.84}$ & $\mathbf{91.84(\pm0.33)}$ & $\mathbf{14.37}$ & $\mathbf{91.81(\pm0.35)}$ & $\mathbf{7.66}$ & $\mathbf{91.85(\pm0.42)}$ & $\mathbf{3.82}$ & $\mathbf{91.84(\pm0.27)}$ & $\mathbf{2.24}$\\   \midrule
  
      \multirow{4}{*}{$\Recall=0.3$}& \ReferenceMethod & $76.13(\pm2.99)$ & $5.02$  & $80.11(\pm2.38)$ & $5.66$ & $81.14(\pm1.64)$ & $4.18$ & $80.74(\pm1.30)$ & $2.69$ & $72.12(\pm4.71)$  & $4.26$ \\  
      & \RandomNoise & $\mathbf{90.97(\pm0.31)}$ & $40.84$ & $\mathbf{90.85(\pm0.24)}$ & $41.92$ & $90.95(\pm0.31)$ & $40.31$ & $90.90(\pm0.37)$ & $41.99$ & $\mathbf{90.99(\pm0.28)}$ & $40.52$\\
      & \FriendlyNoise & $90.48(\pm0.27)$ & $39.25$ & $90.49(\pm0.38)$ & $41.73$ & $90.44(\pm0.35)$ & $38.45$ & $90.31(\pm0.47)$ & $38.81$ & $90.53(\pm0.31)$ & $41.32$\\
      & \sysName & $\mathbf{91.94(\pm0.30)}$ & $\mathbf{31.94}$ & $\mathbf{91.77(\pm0.41)}$ & $\mathbf{21.79}$ & $\mathbf{92.00(\pm0.31)}$ & $\mathbf{19.78}$ & $\mathbf{92.01(\pm0.27)}$ & $\mathbf{5.70}$ &$\mathbf{91.94(\pm0.34)}$ & $\mathbf{2.59}$ \\ 
  
      \bottomrule
      \end{tabular}
      }}
  
      \caption{Experimental results of purification against \emph{backdoor attack} on CIFAR-10 where simulated detectors parameterized by precision
        $\Precision$ and recall $\Recall$ were applied. \GassianVariant
        and \PerturbationMagnitude in baselines were searched such that
        their \Accuracy is as close as possible to \Accuracy of
        \sysName. We bold the numbers following the same
        rule as \tblref{table:poisoning:human:cifar10}.}
      \label{table:backdoor:human:cifar10}
  \end{table*}

  \begin{table*} [ht!]
      \centering
      \vspace{0.1in}
      {\resizebox{\textwidth}{!}{
          \begin{tabular}{@{}lr@{\hspace{0.3em}}rr@{\hspace{0.3em}}rr@{\hspace{0.3em}}rr@{\hspace{0.3em}}rr@{\hspace{0.3em}}rr@{\hspace{0.3em}}r@{}}
          \toprule
              \multicolumn{1}{c}{\multirow{2}{*}{}} & \multicolumn{2}{c}{$\NumThreshold=500$} & \multicolumn{2}{c}{$\NumThreshold=1000$} & \multicolumn{2}{c}{$\NumThreshold=2000$}  & \multicolumn{2}{c}{$\NumThreshold=3000$}  & \multicolumn{2}{c}{$\NumThreshold=4000$} \\
                 & \multicolumn{1}{c}{\Accuracy \%} & \multicolumn{1}{c}{\AttackSuccessRate \%} & \multicolumn{1}{c}{\Accuracy \%} & \multicolumn{1}{c}{\AttackSuccessRate \%} & \multicolumn{1}{c}{\Accuracy \%} & \multicolumn{1}{c}{\AttackSuccessRate \%}  & \multicolumn{1}{c}{\Accuracy \%} & \multicolumn{1}{c}{\AttackSuccessRate \%}  & \multicolumn{1}{c}{\Accuracy \%} & \multicolumn{1}{c}{\AttackSuccessRate \%}  \\
          \midrule
          \ReferenceMethod & $52.11(\pm2.08)$ & $6.26$ & $66.38(\pm1.5)$ & $4.68$ & $77.8(\pm0.69)$ & $4.68$ & $84.24(\pm0.50)$ & $4.34$ & $89.02(\pm0.32)$ & $13.38$ \\
          \RandomNoise & $\mathbf{86.55(\pm0.33)}$ & $\mathbf{4.19}$ & $\mathbf{86.74(\pm0.29)}$ & $\mathbf{6.30}$ & $\mathbf{88.46(\pm0.43)}$ & $25.12$ & $\mathbf{89.08(\pm0.31)}$ & $30.56$ & $\mathbf{90.36(\pm0.35)}$ & $39.08$ \\
          \FriendlyNoise & $\mathbf{86.35(\pm0.61)}$ & $7.51$ & $\mathbf{86.36(\pm0.57)}$ & $8.18$ & $\mathbf{87.76(\pm0.45)}$ & $21.83$ & $\mathbf{89.13(\pm0.40)}$ & $38.34$ & $\mathbf{90.74(\pm0.27)}$ & $57.39$ \\
          \sysName & $\mathbf{86.26(\pm0.38)}$ & $\mathbf{4.95}$ & $\mathbf{86.75(\pm0.56)}$ & $\mathbf{5.15}$ & $\mathbf{88.19(\pm0.38)}$ & $\mathbf{5.07}$ & $\mathbf{89.60(\pm0.31)}$ & $\mathbf{4.99}$ & $\mathbf{90.81(\pm0.35)}$ & $\mathbf{8.44}$ \\
          \bottomrule
          \end{tabular}
          }}
      
          \caption{Experimental results of purification against \emph{backdoor attack} on CIFAR-10 where \knnParameter-NN(\NumThreshold) detectors were
            applied. The precision and recall of such detectors are shown in
            \tblref{table:backdoor:knn-kappa:cifar10:parameters} in
            \appref{sec:app:knn}. We bold the numbers following the same
            rule as \tblref{table:poisoning:human:cifar10}.}
          \label{table:backdoor:knn-kappa:cifar10}
      \end{table*}

\begin{table}[!t]
\centering
\vspace{0.1in}
\resizebox{0.476\textwidth}{!}{
    \begin{tabular}{@{}lr@{\hspace{1.5em}}rr@{\hspace{1.5em}}rr@{\hspace{1.5em}}r@{}}
    \toprule
    \multicolumn{1}{c}{\multirow{2}{*}{}} & \multicolumn{2}{c}{Targeted} & \multicolumn{2}{c}{Backdoor} \\
     & \multicolumn{1}{c}{\Accuracy \%} & \multicolumn{1}{c}{\AttackSuccessRate} & \multicolumn{1}{c}{\Accuracy \%} & \multicolumn{1}{c}{\AttackSuccessRate \%}  \\
    \midrule
    \knnParameter-NN & $86.36(\pm0.44)$ & $3/20$ & $86.30(\pm0.42)$ & $9.67$\\
    \FriendlyNoiseB & $88.49(\pm0.30)$ & $0/20$ & $87.77(\pm0.42)$ & $12.87$\\
    \EPIC & $88.42(\pm0.40)$ & $5/20$ & $88.43(\pm0.36)$ & $16.12$ \\
    \ASD & $78.75(\pm3.46)$ & $2/20$ & $79.03(\pm2.01)$ & $17.56$ \\
    \CBD & $92.84(\pm0.66)$ & $17/20$ & $92.78(\pm0.73)$ & $86.70$ \\
    \sysName & $91.44(\pm0.42)$ & $0/20$ & $91.30(\pm0.37)$ & $2.76$ \\
    \bottomrule
    \end{tabular}
    }
    \caption{Comparison between \sysName and prevention baselines 
    on CIFAR-10. The results of \sysName came from the setting where a simulated detector with
    $\Precision=1.0$ and $\Recall=0.1$ was applied.}
    \label{table:poisoning:baselines}
\end{table}


\myparagraph{Results with purification} There exists a trade-off
between $\Accuracy$ and $\AttackSuccessRate$, in that a purification
method can achieve lower $\AttackSuccessRate$ by adding more
perturbation (e.g., increasing $\GassianVariant$ and
$\PerturbationMagnitude$ in the baselines) into the training data but
that will decrease $\Accuracy$ of the trained model.  Therefore, to
provide a fair comparison, we explored $\GassianVariant \in [16,
\infty)$ in \RandomNoise{\GassianVariant} and $\PerturbationMagnitude \in
[16, \infty)$ in \FriendlyNoise{\PerturbationMagnitude} such that their \Accuracy
was as close as possible to \Accuracy of \sysName.  Then, we compared
their $\AttackSuccessRate$, i.e., a lower $\AttackSuccessRate$
indicated better performance when \Accuracy was comparable to others.

\tblref{table:poisoning:human:cifar10} and \tblref{table:backdoor:human:cifar10} show the results of \sysName and
purification baselines (with searched $\GassianVariant$ or
$\PerturbationMagnitude$) against \emph{targeted attack} and \emph{backdoor attack}
on CIFAR-10 when simulated detectors
with different precisions and recalls were used. First, \sysName
achieved the lowest \AttackSuccessRate when achieving comparable
\Accuracy with the baselines in most cases. In other words, when
achieving comparable utility for the purified training data, the
models learned based on the training data purified by \sysName were
less likely to predict the target inputs or inputs with the 
attacker-chosen trigger as target labels. Second, an
overall trend is that \sysName was better at eliminating the hidden
properties from the learned models when $\Precision$ was larger or
$\Recall$ was smaller. For instance, in the defense against targeted attack, when
$\Precision=0.991$ and $\Recall=0.3$, \AttackSuccessRate of \sysName
is 7/20; when $\Recall$ reduced to 0.1, \AttackSuccessRate reduced to
2/20; and when $\Precision$ further increased to 0.998,
\AttackSuccessRate further reduced to 0/20. This is because the
reference dataset included less manipulated inputs when $\Precision$
was larger or $\Recall$ was smaller, making \sysName more likely to
eliminate the hidden properties. We also explored a simulated detector
with a smaller recall $\Recall < 0.1$ to select reference data (see
\tblref{table:poisoning:human:cifar10:extra} in
\appref{sec:app:extra_results:poisoning}), which further demonstrated that
\sysName outperformed the baselines even when a small amount of
reference data was available. The results of \sysName and
purification baselines on TinyImageNet shown in \tblref{table:poisoning:human:tiny} and \tblref{table:backdoor:human:tiny}
in \appref{sec:app:extra_results:poisoning} also demonstrated the effectiveness and advantages of our proposed method 
against targeted attacks and backdoor attacks.

\tblref{table:poisoning:knn-kappa:cifar10} and \tblref{table:backdoor:knn-kappa:cifar10} show the results when the
\knnParameter-NN($\NumThreshold$) detector with different \NumThreshold was used to
detect the reference data. \sysName outperformed the baselines 
against backdoor attacks (i.e., \sysName achieved lower \AttackSuccessRate when
achieving comparable \Accuracy) and was comparable with them
against targeted attacks  (i.e., achieving comparable \Accuracy and \AttackSuccessRate)
in most cases. \sysName did not outperform the baselines in the 
defense against targeted attacks
because the precision of the \knnParameter-NN($\NumThreshold$) detector was low and
even lower than a random detector (see
\tblref{table:poisoning:knn-kappa:cifar10:parameters} in
\appref{sec:app:knn}). The results on TinyImageNet when the \knnParameter-NN($\NumThreshold$) detector was
applied are shown in \tblref{table:poisoning:knn-kappa:tiny} and \tblref{table:backdoor:knn-kappa:tiny}
in \appref{sec:app:extra_results:poisoning}. These results demonstrated that 
\sysName outperforms the purification baselines on TinyImageNet.

\myparagraph{Comparison with prevention baselines} The comparison
between our proposed method and those prevention baselines is shown in
\tblref{table:poisoning:baselines}.\footnote{Yang, et
al.~\cite{yang2022:not} showed that \EPIC achieved an
\AttackSuccessRate of $0/20$ and an \Accuracy of $ 89.07\%$ against
targeted data poisoning attacks in a setting similar to ours.} From the
results in \tblref{table:poisoning:baselines}, our proposed method
outperforms the state-of-the-art methods like \knnParameter-NN,
\FriendlyNoiseB, and \EPIC against both clean-label targeted attacks
and clean-label backdoor attacks when a good detector of reference
data is selected.  The two recently proposed defenses, \ASD and \CBD,
were not effective, though they have been demonstrated to be powerful
in the defense against dirty-label backdoor
attacks~\cite{gao2023:backdoor,zhang2023:backdoor}.


 \section{As an Attack on Data Tracing} \label{sec:tracing}

\subsection{Radioactive Data} \label{sec:tracing:prelim}

Generally, data tracing is a technique used to modify (``mark'') a
subset of training inputs such that a data contributor can detect
whether her marked training inputs were used to train a given
model. The modifications are carefully crafted such that the
parameters of a model trained on the marked training data have
property that can be detected by the data contributor. We take
\textit{Radioactive Data}~\cite{sablayrolles2020:radioactive}, the state-of-the-art data tracing
method, as an example to illustrate the key idea of data
tracing.

\myparagraph{Manipulating data} Radioactive Data assumes the data
contributor marks a subset of the training inputs based on a feature
extractor $\Network{{\markedTag}}$ called the \emph{marking feature
extractor}. Specifically, the data contributor selects a random unit
vector $\Mark{\classIdx}$ for each class
$\classIdx=1,2,\cdots,\NumClasses$. The unit vector $\Mark{\classIdx}$
has the same number of dimensions as the feature vector output by
$\Network{{\markedTag}}$. The unit vectors
$\{\Mark{\classIdx}\}_{\classIdx=1}^{\NumClasses}$ are called
``marks''. These marks are used to guide how to modify the training
inputs. In particular, for each selected training input $\Image$ to be
marked in class $\classIdx$, Radioactive Data aims to add a
modification $\ManipulationPerturbation$ to $\Image$ such that 1) the feature
vector of $\Image + \ManipulationPerturbation$, i.e.,
$\Network{{\markedTag}}(\Image + \ManipulationPerturbation)$, is similar to
the mark $\Mark{\classIdx}$; 2) $\Image + \ManipulationPerturbation$ and
$\Image$ have similar feature vectors; 3) $\Image + \ManipulationPerturbation$
is close to $\Image$ with respect to their pixel values; and 4)
$\infinityNorm{\ManipulationPerturbation}\leq \TracingHyperparameter$.  (2),
(3), and (4) aim to preserve the utility of the marked training
inputs, while (1) aims to make a model trained on the marked training
data have a property that the data contributor can detect with
a statistical guarantee, which we discuss next.

\myparagraph{Detecting property} Suppose a model is trained on a
training dataset, a subset of which was marked by Radioactive Data. To
illustrate the data-contributor-detectable property of the model and
how to detect it, we assume for simplicity that the model uses the
same marking feature extractor and adds a linear layer on top of it.
When a model does not use the marking feature extractor (this is the
case in our experiments), the data contributor can decompose it into a
feature extractor and a linear layer, align the two feature extractors
by applying a transformation of the second feature extractor, and then
detect the property after alignment~\cite{sablayrolles2020:radioactive}.

We denote the parameters of the added linear layer for class
${\classIdx}$ as $\NetworkParams_{\classIdx}$. If the model does
\emph{not} use the marked data for training, then
$\NetworkParams_{\classIdx}$ is not statistically related to the mark
$\Mark{\classIdx}$. Since a mark $\Mark{\classIdx}$ is picked
uniformly at random, the cosine similarity between
$\NetworkParams_{\classIdx}$ and $\Mark{\classIdx}$ follows a prior
distribution, in particular an incomplete beta distribution~\cite{iscen2017:memory}.
When the model uses the marked data for training, then
$\NetworkParams_{\classIdx}$ is statistically related to the mark
$\Mark{\classIdx}$. In particular, $\NetworkParams_{\classIdx}$ would
be similar to $\Mark{\classIdx}$ since many training inputs (i.e., the
marked ones) in class $\classIdx$ have feature vectors
similar to $\Mark{\classIdx}$. As a result, the cosine similarity
between $\NetworkParams_{\classIdx}$ and $\Mark{\classIdx}$ would
significantly deviate from the prior incomplete beta distribution.

Formally, Radioactive Data uses hypothesis testing to detect whether
this cosine similarity significantly deviates from the prior
distribution and combines hypothesis testing for the $\NumClasses$
classes to obtain a result. If the testing favors the hypothesis that
the cosine similarity is not from the prior incomplete beta
distribution under a significance level (e.g., a $\PValue$-value with
$\PValue<0.05$ used in both Radioactive Data and our experiments),
then the data contributor concludes that her marked data was used to
train the model. When applying our \sysName to attack data tracing,
the goal is to purify the marked training data such that the data
contributor cannot detect the use of her data, i.e., the
$\PValue$-value in the hypothesis testing is no smaller than 0.05.

\subsection{Experimental Setup}  \label{sec:tracing:setup}

\myparagraph{Datasets} 
demonstrated by previous work~\cite{tekgul2022:effectiveness},
\textit{Radioactive Data} is less effective in the cases where the
dataset marked by it has a small number of classes, e.g.,
CIFAR-10. Therefore, we used two visual benchmarks:
CIFAR-100~\cite{krizhevsky2009:learning} and
TinyImageNet~\cite{Lle2015:TinyIV}.
\begin{itemize}[nosep,leftmargin=1em,labelwidth=*,align=left]
\item \textbf{CIFAR-100.} CIFAR-100 is a dataset containing $60{,}000$
  $32\times32\times3$ images partitioned into $\NumClasses=100$
  classes in which $50{,}000$ images are training samples and
  $10{,}000$ are testing samples.

\item \textbf{TinyImageNet.} TinyImageNet is a dataset containing
  $64\times64\times3$ images partitioned into $\NumClasses=200$
  classes. In each class, there are $500$ training images and $50$
  validation images.  We sampled $50$ classes (see
  \appref{sec:app:sample}) and used the data from these $50$ classes
  in our experiments. As such, there are $25{,}000$ training images
  and $2{,}500$ testing images. \emph{Due to the space limit,
  we displayed most of results on TinyImageNet in \appref{sec:app:extra_results:tracing}}.
\end{itemize}

\myparagraph{Marking setting}
We setup the marking setting of Radioactive Data represented by
$\{\Dataset{\trainTag},
\{\Mark{\classIdx}\}_{\classIdx=1}^{\NumClasses}\}$ in one experiment
as follows: First, we randomly sampled $\NumClasses$ unit vectors as
$\{\Mark{\classIdx}\}_{\classIdx=1}^{\NumClasses}$.  Next, we
uniformly sampled $10\%$ of training samples (i.e., $50$ samples
per class in CIFAR-100 and $50$ samples per class in TinyImageNet) to
be marked by Radioactive Data. As such, we constructed the marked
training dataset $\Dataset{\trainTag}$.  Following the implementation
of  previous
work\footnote{\url{https://github.com/facebookresearch/radioactive\_data}},
we constructed $20$ sets of $\{\Dataset{\trainTag},
\{\Mark{\classIdx}\}_{\classIdx=1}^{\NumClasses}\}$ for CIFAR-100 and
another $20$ sets of $\{\Dataset{\trainTag},
\{\Mark{\classIdx}\}_{\classIdx=1}^{\NumClasses}\}$ for TinyImageNet.

\myparagraph{Selecting reference data}  
We used the simulated detector parameterized by precision
$\Precision$ and recall $\Recall$ and adapted $\knnParameter$-NN detector to
obtain the reference dataset $\Dataset{\referenceTag}$. The simulated
detector and adapted $\knnParameter$-NN detector are described in
\secref{sec:poisoning:setup}. Note that 10\% of the training data is marked. 
Therefore, we consider the precision of the simulated detector to be $\Precision>0.90$ 
so that it is better than the random detector that selects the reference data from 
the marked training dataset uniformly at random.

\myparagraph{Purification setting}   
In our experiments, we used ResNet-18 pretrained on ImageNet as the
feature extractor $\Network$.  We set 
$\GradientPenaltyCoefficient = 10$;  $\BatchSize = 256$;
 $\LearningRate{\Critic} = 0.001$;
$\LearningRate{\PurifiedPerturbation} = 2.0$ for 
CIFAR-100 and $\LearningRate{\PurifiedPerturbation} = 10.0$ for
 TinyImageNet;  $\SelectionRatio =
5\%$;  $\Amplification=2.0$; and the interval of
checkpoints $\CheckingPoint = 1$.  We set the regularization parameter
$\PurificationHyperparameter$ as $0$ and stopped alternating between
the two steps in \algref{ag:purification} when the number of
iterations reached $30$ to bound the perturbations in our experiments.

\myparagraph{Setting to train models}
All experiments are evaluated on ResNet-18 models trained on
normalized data applying default data augmentation by a SGD optimizer
for $40$ epochs.  The batch size was $128$, and the initial learning
rate was $0.1$ decayed by a factor of $0.1$ when the number of epochs
reached $15$, $25$, or $35$.

\subsection{Baselines}  \label{sec:tracing:baseline}

We considered \ReferenceMethod, \RandomNoise{\GassianVariant}, and
\FriendlyNoise{\PerturbationMagnitude}, introduced in
\secref{sec:poisoning:baseline} as the baselines, where $\GassianVariant \in [16, \infty)$ and $\PerturbationMagnitude \in [16, \infty)$.

\begin{table} [!t]
\centering
\vspace{0.1in}
    \begin{tabular}{@{}lr@{\hspace{0.9em}}rr@{\hspace{0.9em}}r@{}}
    \toprule
         \multicolumn{2}{c}{CIFAR-100} & \multicolumn{2}{c}{TinyImageNet} \\
     \multicolumn{1}{c}{\Accuracy \%} & \multicolumn{1}{c}{$\PValue < 0.05$} & \multicolumn{1}{c}{\Accuracy \%} & \multicolumn{1}{c}{$\PValue < 0.05$}  \\
    \midrule
     $70.98(\pm0.28)$ & $20/20$ & $60.64(\pm1.05)$ & $19/20$\\
    \bottomrule
    \end{tabular}
    \caption{Average $\Accuracy$ and detection results of the models
      trained on the marked training datasets. The numbers in the
      parenthesis are standard deviations among the 20 experiment
      trials. $\PValue<0.05$ means that the data contributor
      successfully detected her marked data was used in model
      training among the 20 experiment trials.  }
    \label{table:tracing:performance}
\end{table}


\begin{table*} [ht!]
  \centering
  \vspace{0.1in}
  {\resizebox{\textwidth}{!}{
      \begin{tabular}{@{}lr@{\hspace{1.5em}}rr@{\hspace{1.5em}}rr@{\hspace{1.5em}}rr@{\hspace{1.5em}}rr@{\hspace{1.5em}}rr@{\hspace{1.5em}}rr@{\hspace{1.5em}}r@{}}
      \toprule
          \multicolumn{1}{c}{\multirow{2}{*}{}} & \multicolumn{1}{c}{} & \multicolumn{2}{c}{$\Precision=0.92$} & \multicolumn{2}{c}{$\Precision=0.94$} & \multicolumn{2}{c}{$\Precision=0.96$}  & \multicolumn{2}{c}{$\Precision=0.98$}  & \multicolumn{2}{c}{$\Precision=1.0$} \\
            & & \multicolumn{1}{c}{\Accuracy \%} & \multicolumn{1}{c}{$\PValue < 0.05$} & \multicolumn{1}{c}{\Accuracy \%} & \multicolumn{1}{c}{$\PValue < 0.05$} & \multicolumn{1}{c}{\Accuracy \%} & \multicolumn{1}{c}{$\PValue < 0.05$}  & \multicolumn{1}{c}{\Accuracy \%} & \multicolumn{1}{c}{$\PValue < 0.05$}  & \multicolumn{1}{c}{\Accuracy \%} & \multicolumn{1}{c}{$\PValue < 0.05$}  \\
      \midrule
      
      \multirow{4}{*}{$\Recall=0.1$}& \ReferenceMethod &$29.29(\pm0.99)$ & $1/20$ & $29.04(\pm0.70)$ & $1/20$ & $28.82(\pm0.70)$ & $3/20$ & $27.04(\pm0.88)$ & $1/20$ & $26.92(\pm0.78)$ & $1/20$  \\  
      & \RandomNoise &$\mathbf{67.31(\pm0.32)}$ & $10/20$ & $\mathbf{67.27(\pm0.43)}$ & $12/20$ & $\mathbf{67.20(\pm0.28)}$ & $8/20$ & $\mathbf{67.21(\pm0.28)}$ & $12/20$ & $\mathbf{67.29(\pm0.36)}$ & $10/20$  \\
      & \FriendlyNoise &$\mathbf{67.27(\pm0.31)}$ & $8/20$ & $\mathbf{67.26(\pm0.26)}$ & $13/20$ & $\mathbf{67.11(\pm0.29)}$ & $9/20$ & $\mathbf{67.20(\pm0.30)}$ & $10/20$ & $\mathbf{67.17(\pm0.36)}$ & $12/20$  \\
      & \sysName &$\mathbf{67.52(\pm0.22)}$ & $\mathbf{4/20}$ & $\mathbf{67.56(\pm0.27)}$ & $\mathbf{3/20}$ & $\mathbf{67.57(\pm0.30)}$ & $\mathbf{2/20}$ & $\mathbf{67.61(\pm0.45)}$ & $\mathbf{4/20}$ & $\mathbf{67.63(\pm0.28)}$ & $\mathbf{4/20}$  \\\midrule
  
      \multirow{4}{*}{$\Recall=0.2$}& \ReferenceMethod & $44.71(\pm0.69)$ & $1/20$ & $44.21(\pm0.74)$ & $2/20$ & $43.85(\pm0.92)$ & $0/20$ & $41.28(\pm1.22)$ & $0/20$ & $43.48(\pm0.81)$ & $0/20$ \\ 
      & \RandomNoise &$\mathbf{67.84(\pm0.32)}$ & $12/20$ & $\mathbf{67.76(\pm0.44)}$ & $7/20$ & $\mathbf{67.77(\pm0.32)}$ & $10/20$ & $\mathbf{67.97(\pm0.19)}$ & $13/20$ & $\mathbf{67.64(\pm0.31)}$ & $10/20$  \\
      & \FriendlyNoise &$\mathbf{67.87(\pm0.38)}$ & $14/20$ & $\mathbf{67.73(\pm0.29)}$ & $8/20$ & $\mathbf{67.59(\pm0.27)}$ & $12/20$ & $\mathbf{67.64(\pm0.32)}$ & $14/20$ & $\mathbf{67.84(\pm0.34)}$ & $11/20$  \\
      & \sysName &$\mathbf{68.33(\pm0.37)}$ & $\mathbf{6/20}$ & $\mathbf{68.37(\pm0.29)}$ & $\mathbf{6/20}$ & $\mathbf{68.55(\pm0.32)}$ & $\mathbf{2/20}$ & $\mathbf{68.55(\pm0.23)}$ & $\mathbf{3/20}$ & $\mathbf{68.53(\pm0.34)}$ & $\mathbf{3/20}$  \\ \midrule
  
      \multirow{4}{*}{$\Recall=0.3$}& \ReferenceMethod & $51.45(\pm0.52)$ & $1/20$ & $53.14(\pm0.63)$ & $2/20$ & $52.63(\pm0.59)$ & $2/20$ & $45.08(\pm2.02)$ & $2/20$ & $52.32(\pm0.66)$ & $1/20$ \\ 
      & \RandomNoise &$\mathbf{68.13(\pm0.27)}$ & $12/20$ & $\mathbf{68.13(\pm0.27)}$ & $11/20$ & $\mathbf{68.36(\pm0.28)}$ & $14/20$ & $\mathbf{68.27(\pm0.32)}$ & $14/20$ & $\mathbf{68.21(\pm0.25)}$ & $10/20$  \\
      & \FriendlyNoise &$\mathbf{68.10(\pm0.27)}$ & $14/20$ & $\mathbf{68.18(\pm0.35)}$ & $15/20$ & $\mathbf{68.12(\pm0.28)}$ & $14/20$ & $\mathbf{68.17(\pm0.26)}$ & $12/20$ & $68.07(\pm0.29)$ & $10/20$  \\
      & \sysName &$\mathbf{68.87(\pm0.34)}$ & $\mathbf{10/20}$ & $\mathbf{68.90(\pm0.24)}$ & $\mathbf{8/20}$ & $\mathbf{69.00(\pm0.36)}$ & $\mathbf{6/20}$ & $\mathbf{69.07(\pm0.41)}$ & $\mathbf{2/20}$ & $\mathbf{69.18(\pm0.31)}$ & $\mathbf{4/20}$  \\
  
      \bottomrule
      \end{tabular}
      }}
  
      \caption{Experimental results of purification against \emph{data tracing}
        on CIFAR-100 where
        simulated detectors parameterized by precision $\Precision$ and
        recall $\Recall$ were applied. \GassianVariant and
        \PerturbationMagnitude in baselines were searched such that
        their \Accuracy are as close as possible to \Accuracy of
        \sysName. We bold the numbers following the same rule as
        \tblref{table:poisoning:human:cifar10}.
      }
      \label{table:tracing:human:cifar100}
  \end{table*}

  \begin{table*} [ht!]
  \centering
  \vspace{0.1in}
  {\resizebox{\textwidth}{!}{
      \begin{tabular}{@{}lr@{\hspace{0.3em}}rr@{\hspace{0.3em}}rr@{\hspace{0.3em}}rr@{\hspace{0.3em}}rr@{\hspace{0.3em}}rr@{\hspace{0.3em}}r@{}}
          \toprule
          \multicolumn{1}{c}{\multirow{2}{*}{}} & \multicolumn{2}{c}{$\NumThreshold=50$} & \multicolumn{2}{c}{$\NumThreshold=100$} & \multicolumn{2}{c}{$\NumThreshold=150$}  & \multicolumn{2}{c}{$\NumThreshold=200$}  & \multicolumn{2}{c}{$\NumThreshold=250$} \\
             & \multicolumn{1}{c}{\Accuracy \%} & \multicolumn{1}{c}{$\PValue < 0.05$} & \multicolumn{1}{c}{\Accuracy \%} & \multicolumn{1}{c}{$\PValue < 0.05$} & \multicolumn{1}{c}{\Accuracy \%} & \multicolumn{1}{c}{$\PValue < 0.05$}  & \multicolumn{1}{c}{\Accuracy \%} & \multicolumn{1}{c}{$\PValue < 0.05$}  & \multicolumn{1}{c}{\Accuracy \%} & \multicolumn{1}{c}{$\PValue < 0.05$}  \\
      \midrule
      \ReferenceMethod & $28.10(\pm1.23)$ & $1/20$ & $42.61(\pm0.96)$ & $0/20$ & $50.09(\pm0.39)$ & $1/20$ & $55.39(\pm0.32)$ & $0/20$ & $59.43(\pm0.35)$ & $1/20$ \\
      \RandomNoise & $\mathbf{65.42(\pm0.27)}$ & $5/20$ & $\mathbf{66.60(\pm0.26)}$ & $8/20$ & $\mathbf{66.68(\pm0.22)}$ & $13/20$ & ${66.91(\pm0.23)}$ & $13/20$ & $67.49(\pm0.39)$ & $13/20$ \\
      \FriendlyNoise & $\mathbf{65.30(\pm0.34)}$ & $8/20$ & $\mathbf{66.59(\pm0.28)}$ & $10/20$ & $\mathbf{66.70(\pm0.27)}$ & $10/20$ & $66.78(\pm0.31)$ & $12/20$ & $67.49(\pm0.28)$ & $13/20$ \\
      \sysName & $\mathbf{65.54(\pm0.32)}$ & $\mathbf{2/20}$ & $\mathbf{66.38(\pm0.33)}$ & $\mathbf{2/20}$ & $\mathbf{67.34(\pm0.28)}$ & $\mathbf{5/20}$ & $\mathbf{68.02(\pm0.23)}$ & $\mathbf{7/20}$ & $\mathbf{68.80(\pm0.25)}$ & $\mathbf{6/20}$ \\
      \bottomrule
      \end{tabular}
      }}
  
      \caption{Experimental results of purification against \emph{data tracing}
        on CIFAR-100 where
        \knnParameter-NN(\NumThreshold) detectors were applied. The
        precision and recall of such detectors are shown in
        \tblref{table:tracing:knn-kappa:cifar100:parameters} in
        \appref{sec:app:knn}. We bold the numbers following the same
        rule as \tblref{table:poisoning:human:cifar10}.}
      \label{table:purification:knn-kappa:cifar100}
  \vspace{0.1in}
  \end{table*}

\subsection{Experimental Results}   \label{sec:tracing:result}

\myparagraph{Results without purification}
\tblref{table:tracing:performance} shows the results when the
marked training dataset is not purified. $\Accuracy$ is the testing
accuracy of the models trained on the marked training data, and
$\PValue<0.05$ means that the data contributor can detect that her
marked data was used to train a given model. Our results confirm
that Radioactive Data enables a data contributor to successfully
detect unauthorized data use in model training with statistical
guarantees when the marked training data is not purified.

\myparagraph{Results with purification} Like the experiments in
defending against data poisoning attacks, there exists a trade-off in
data purification between the accuracy \Accuracy of the trained model
and the frequency at which the data contributor detects the use of her
data (with significance value $\PValue < 0.05$). Therefore, to compare
\sysName with those baselines fairly, we searched the perturbation
magnitude $\GassianVariant \in [16, \infty)$ and
$\PerturbationMagnitude \in [16, \infty)$ in \RandomNoise and
 \FriendlyNoise such that
their \Accuracy are as close as possible to \Accuracy of
\sysName. \tblref{table:tracing:human:cifar100} shows the results of
different methods on CIFAR-100 when the simulated detector with different
$\Precision$ and $\Recall$ is used to select the reference data.

\sysName outperforms the baselines in most cases. Specifically, when
achieving comparable \Accuracy, \sysName achieves lower frequency of
detection (i.e., $\PValue < 0.05$) in most cases. In other words, when
preserving the data utility to the same extent, \sysName is better at
eliminating the hidden properties from the models learned using the
purified training data.  \sysName also achieves lower frequency of
detection (i.e., more likely to eliminate the hidden properties) when
\Recall of the simulated detector is smaller, especially when
\Precision is small. For instance, when $\Precision=0.92$, the
frequency of detection decreases from 10/20 to 4/20 when \Recall
decreases from 0.3 to 0.1 on CIFAR-100. This is because, given a
\Precision, a smaller \Recall indicates less number of marked training
inputs in the reference dataset. However, this trend is less
noticeable when \Precision is very large (e.g., $\Precision=1.0$),
because the number of marked training inputs in the reference dataset
is always small for different \Recall in such cases. The results
on TinyImageNet shown in \tblref{table:tracing:human:tiny} in \appref{sec:app:extra_results:tracing} also
demonstrated the advantages of our proposed method.

\tblref{table:purification:knn-kappa:cifar100} shows the results on CIFAR-100 when the
reference data is selected using the adapted \knnParameter-NN detector
with different $\NumThreshold$.  \sysName also outperforms the
baselines when such a detector is used.  Specifically, \sysName
achieves lower detection frequency than the baselines when achieving
comparable \Accuracy with them. Moreover, \sysName achieves smaller
detection frequency on CIFAR-100 than on TinyImageNet (shown in \tblref{table:tracing:knn-kappa:tiny}
in \appref{sec:app:extra_results:tracing}) when the
\knnParameter-NN detector uses the same $\NumThreshold$. This is
because the detected reference data is ``cleaner'' on CIFAR-100. For
instance, the models learned on the reference data alone (i.e., the
rows corresponding to $\ReferenceMethod$) have lower detection
frequency on CIFAR-100.


\section{Discussion}  \label{sec:discussion}

\subsection{Adaptive Training-Data manipulation}

\sysName minimizes the distance between the distributions of the
untrusted data and that of the reference data. In response, a data
contributor might constrain the manipulation such that the
distribution of the manipulated data is closer to that of the
unmanipulated data, causing \sysName to minimally perturb the training
inputs since the distribution distance is already small. To test such
a strategy, we considered an adaptive data contributor that tunes
the bound of the manipulation (\PoisonHyperparameter).  A smaller
manipulation bound (\PoisonHyperparameter) makes the distribution of
the manipulated data more similar to the one of the reference data. To
implement such adaptive strategies, we adapted data poisoning
(described in \secref{sec:poisoning}) by changing
\PoisonHyperparameter from $16$ (the default) to $4$, and data tracing
(described in \secref{sec:tracing}) by changing \PoisonHyperparameter
from $10$ (the default) to $2$.  Our experimental results (see
\tblref{table:adaptive} in \appref{app:adaptive}) show that when \PoisonHyperparameter was
tuned down, \sysName was slightly degraded but was still effective to
remove the hidden properties without significantly affecting the
utility of the data. However, at the same time, the decrease of
\PoisonHyperparameter also reduced effectiveness of training-data
manipulation. The study demonstrates the robustness of \sysName to
the adaptive training-data manipulation. Details of this study are shown in \appref{app:adaptive}.

\subsection{Ethical Statement}

\sysName can be used as an attack on data tracing methods such as
\textit{Radioactive Data}. We believe that our work will benefit the
security community in this area by providing a tool to test the
robustness of data tracing methods in the future.

\section{Conclusion} \label{sec:conclusion}

In this work, we propose \sysName, a general framework to purify
manipulated data. We find that, via adding carefully crafted
perturbations to the potentially manipulated training inputs, we can
make the model learned on the purified training data free of hidden
properties intended by the data contributor, while retaining the
accuracy of the learned model.  Given a reference dataset, finding
such perturbations can be formulated as a min-max optimization
problem, which can be approximately solved by a two-step method.
\sysName achieves good performance as long as the detector used to
detect the reference data provides high precision, though it need not
be perfect.  We demonstrated the generality of \sysName using two
representative use cases, namely defending against data poisoning and
data tracing.  Interesting future work includes applying \sysName into other
types of data as well as
developing purification methods with formal guarantees.


\bibliographystyle{IEEEtran}
\bibliography{full, references}

\appendix

\section{Sampled Classes of TinyImageNet}  \label{sec:app:sample}

We used $50$ classes of TinyImageNet in our experiments. The sampled $50$ classes of TinyImageNet are listed in \tblref{table:classes_tinyimagenet}.

\begin{table}[ht!]
\footnotesize
\begin{center}
\resizebox{0.47\textwidth}{!}{
\begin{tabular}{@{}ccc|ccc@{}}
\toprule
Index & Text & Word & Index & Text & Word \\ \midrule
 0 & n01443537  & goldfish     & 25 & n03255030  & dumbbell        \\ 
 1 & n01698640  & American alligator   & 26 & n03400231 & frying pan          \\ 
 2 & n01774384  & black widow   & 27 & n03447447 & gondola       \\ 
 3 & n01882714  & koala    & 28 & n03617480 & kimono       \\ 
 4 & n01945685  & slug     & 29 & n03670208 & limousine      \\ 
 5 & n02002724  & black stork  & 30 & n03770439 & miniskirt      \\ 
 6 & n02085620  & Chihuahua   & 31 & n03837869  & obelisk      \\
 7 & n02106662  & German shepherd   & 32 & n03902125 & pay-phone      \\ 
 8 & n02124075  & Egyptian cat   & 33 & n03976657 & pole     \\ 
 9 & n02165456  & ladybug  & 34 & n03992509 & potter's wheel       \\ 
 10 & n02231487  & walking stick  & 35 & n04070727 & refrigerator       \\ 
 11 & n02279972  & monarch   & 36 & n04133789 & sandal      \\ 
 12 & n02395406  & hog  & 37 & n04251144 & snorkel       \\ 
 13 & n02423022  & gazelle  & 38 & n04275548 & spider web       \\ 
 14 & n02486410  & baboon   & 39 & n04356056 & sunglasses      \\ 
 15 & n02669723  & academic gown   & 40 & n04398044 & teapot      \\ 
 16 & n02788148  & bannister    & 41 & n04465501 & tractor     \\ 
 17 & n02802426  & basketball  &  42 & n04507155 & umbrella       \\ 
 18 & n02815834  & beaker  &  43 & n04560804 & water jug       \\ 
 19 & n02843684  & birdhouse  &  44 & n06596364 & comic book        \\ 
 20 & n02909870  & bucket  & 45 & n07615774 & ice lolly        \\ 
 21 & n02950826  & cannon  & 46 & n07720875 & bell pepper       \\ 
 22 & n02999410  & chain   &  47 & n07753592 & banana      \\ 
 23 & n03085013  & computer keyboard   & 48 & n07875152 & potpie      \\ 
 24 & n03160309  & dam   & 49 & n09256479 & coral reef       \\ 
 \bottomrule 
\end{tabular}
}
\end{center}
\caption{Classes of TinyImageNet used in \secref{sec:poisoning:setup} and \secref{sec:tracing:setup}}
\label{table:classes_tinyimagenet}
\end{table}

\section{\knnParameter-NN(\NumThreshold) Detector}  \label{app:adapted_knn}

In \secref{sec:poisoning} and \secref{sec:tracing}, we 
consider a real-world detector to select reference data. Such a real-world detector
is an adapted $\knnParameter$-NN denoted as $\knnParameter$-NN($\NumThreshold$), modified
from the state-of-the-art detection method, $\knnParameter$-NN~\cite{peri2020:deep}. Here, we introduce both the state-of-the-art detection
method, $\knnParameter$-NN, and its adapted version, $\knnParameter$-NN($\NumThreshold$).

\myparagraph{$\knnParameter$-NN detection~\cite{peri2020:deep}}
The $\knnParameter$-NN detection~\cite{peri2020:deep} works as follows: 
\begin{itemize}[nosep,leftmargin=1em,labelwidth=*,align=left]
    \item 1) A feature extractor (e.g., the same feature extractor used for
    purification) is used to extract features of the inputs in
    $\Dataset{\trainTag}$;
    \item 2) A $\knnParameter$-NN model is trained on
    the extracted features with labels;
    \item 3) For each image in
    $\Dataset{\trainTag}$, if its label is the same as that predicted by
    the trained $\knnParameter$-NN model, it is selected as
    ``unmanipulated'';
    \item 4) A model is trained on the selected ``unmanipulated'' dataset.
\end{itemize}

\myparagraph{$\knnParameter$-NN($\NumThreshold$)}
We modified $\knnParameter$-NN detection as $\knnParameter$-NN($\NumThreshold$) 
by selecting $\NumThreshold$ most confident samples per class as ``unmanipulated''.
$\knnParameter$-NN($\NumThreshold$) works as follows: 
\begin{itemize}[nosep,leftmargin=1em,labelwidth=*,align=left]
    \item 1) A feature extractor (e.g., the same feature extractor used for
    purification) is used to extract features of the inputs in
    $\Dataset{\trainTag}$;
    \item 2) A $\knnParameter$-NN model is trained on
    the extracted features with labels;
    \item 3) For each class, images are ranked according to each
    one's confidence score of being the associated label;
    \item 4) The
    top $\NumThreshold$ samples are selected as ``unmanipulated'' and
    returned as the reference dataset $\Dataset{\referenceTag}$.
\end{itemize}

\section{Baselines and Their Implementation Details} \label{app:baseline}

\myparagraph{Random Noise ($\RandomNoise{\GassianVariant}$)} 
This method constructs perturbations that are sampled from a
Gaussian distribution. $\RandomNoise{\GassianVariant}$ 
adds perturbation into the images from the poisoned training dataset
excluding the reference data (i.e., $\Dataset{\untrustedTag} =
\Dataset{\trainTag} \setminus \Dataset{\referenceTag}$). 
Here $\GassianVariant$ denotes the
standard deviation of a Gaussian distribution. Specifically, when
adding perturbation to an image, this method adds a random noise
sampled from a Gaussian distribution with mean 0 and standard
deviation $\GassianVariant$ to each pixel value of the image. We
explored  $\GassianVariant \in [16, \infty)$.

\myparagraph{Friendly Noise ($\FriendlyNoise{\PerturbationMagnitude}$)~\cite{liu2022:friendly}} 
This method constructs a perturbation that is maximized
within a ball of radius $\PerturbationMagnitude$ in the infinity norm
but that, when added to the data, does not substantially change the
model's output. Specifically, the method first trains a model
$\LinearClassifier{}$ on $\Dataset{\trainTag}$.
Then, for an input
$\Image$, \textit{Friendly Noise} purifies it by adding a perturbation
$\noise$ to it, where $\noise$ is constructed by solving the following
optimization problem: $\min_{\noise: \infinityNorm{\noise}\leq \PerturbationMagnitude} \KLDivergence{\LinearClassifier(\Image + \noise)}{\LinearClassifier(\Image)} - \FriendlynoiseHyperparameter\euclideanNorm{\noise}$,  
where $\PerturbationMagnitude \in \reals$ and
$\FriendlynoiseHyperparameter \in \reals$ are two hyperparameters
controlling the magnitude of the perturbation, and
$\KLDivergence{\LinearClassifier{(\Image +
    \noise)}}{\LinearClassifier(\Image)}$ denotes the KL-divergence
between $\LinearClassifier{(\Image + \noise)}$ and
$\LinearClassifier(\Image)$, which are model outputs for $\Image +
\noise$ and $\Image$, respectively.

We denote the \textit{Friendly Noise} method as
$\FriendlyNoise{\PerturbationMagnitude}$ that purifies the inputs in
$\Dataset{\untrustedTag}$.
We followed the previous
work\footnote{\url{https://github.com/tianyu139/friendly-noise}}
in the implementation of $\FriendlyNoise{\PerturbationMagnitude}$. We set
$\FriendlynoiseHyperparameter = 1$  and
explored different options of $\PerturbationMagnitude \in [16,
\infty)$.

\myparagraph{\knnParameter-NN~\cite{peri2020:deep}}
See the description of \knnParameter-NN detection~\cite{peri2020:deep} in \appref{app:adapted_knn}.
In the implementation of \knnParameter-NN detection, we set \knnParameter as $5000$ for
CIFAR-10 and $500$ for CIFAR-100 or TinyImageNet.

\myparagraph{\FriendlyNoiseB~\cite{liu2022:friendly}}
\FriendlyNoiseB adds friendly noises to all the trainining data before model training and introduces
Bernoulli noises $\noise \in \{-\BernoulliNoiseParameter, \BernoulliNoiseParameter\}$ to 
augment training samples during model training. As such, \FriendlyNoiseB is a 
combination of purification and prevention. \FriendlyNoiseB is considered as the state-of-the-art defense method against
clean-label data poisoning attacks~\cite{geiping2020:witches,souri2022:sleeper}. The construction of friendly noise is described in
the method $\FriendlyNoise{\PerturbationMagnitude}$. The Bernoulli noises are sampled from
a Bernoulli distribution and are added to each training batch in model training. We followed
the previous work~\cite{liu2022:friendly} to implement this method, 
where we set $\PerturbationMagnitude=16$, $\BernoulliNoiseParameter=16$, and the other hyperparameters to the default values.

\myparagraph{\EPIC~\cite{yang2022:not}}
\EPIC is a robust learning algorithm where the training set is dynamically updated by removing samples located 
in low-density gradient space during the learning process. Specifically, the basic idea of \EPIC is designed as follows:
after $\EPICWarmupParameter$ epochs of model training, for every \EPICEpochParameter epochs, it searches for $\EPICSubsetSizeParameter \times \setSize{\Dataset{\trainTag}}$ most centrally located 
training samples, known as \emph{medoids}, in the gradient domain, assigns the other samples into these medoids, and then removes
the medoids that no other sample is assigned to, from the training set. As such, the training set keeps being updated during model training
where the effective poisoned data is removed. \EPIC has been demonstrated to be effective against \textit{Gradient Matching}-based
clean-label targeted poisoning attack and clean-label backdoor attack~\cite{yang2022:not,liu2022:friendly}. We followed the previous 
work\footnote{\url{https://github.com/YuYang0901/EPIC/tree/master}} to implement \EPIC. As suggested by the authors of \EPIC, we set
$\EPICWarmupParameter = 1$, $\EPICEpochParameter=5$, and $\EPICSubsetSizeParameter=0.2$. We set the other hyperparameters to the default values.

\myparagraph{\ASD~\cite{gao2023:backdoor}}
\ASD is a robust learning algorithm that dynamically splits samples from the training set to learn. It includes $3$ steps:
at the first step when the number of epochs is less than $\ASDEpochParameter{1}$, starting from $10$ clean samples per class as the clean set,
for every $5$ epochs, it moves $10$ samples with the lowest symmetric cross-entropy loss per class from the remaining training set to the clean set and leaves
the remaining as the polluted set;
at the second step when the number of epochs is less than $\ASDEpochParameter{2}$, it adds $50\%$ samples of the entire training set with the lowest loss
into the clean set and leaves the remaining as the polluted set; at the third step when the number of epochs is less than $\ASDEpochParameter{3}$,
it applies a meta-split method to detect the hard clean samples from the polluted set. While the clean set and polluted set are
dynamically changed, the model is updated by being supervisedly trained on the clean set and semi-supervisedly trained on the polluted set. \ASD has been demonstrated
as a powerful defense against dirty-label backdoor attacks. We followed the previous work\footnote{\url{https://github.com/KuofengGao/ASD/tree/main}} to implement it,
where $\ASDEpochParameter{1}=60$, $\ASDEpochParameter{2}=90$, and $\ASDEpochParameter{3}=120$. We set the other hyperparameters to the default values.

\myparagraph{\CBD~\cite{zhang2023:backdoor}}
\CBD is a learning algorithm inspired by causal inference~\cite{peters2017:elements}. It firstly trained a backdoored model on the poisoned dataset for $5$ epochs.
Then a clean model is trained for $100$ epochs by reweighting the training samples such that the clean model is independent of the backdoored model in the hidden space.
\CBD is a recently proposed powerful defense against dirty-label backdoor attacks. We followed the previous work\footnote{\url{https://github.com/zaixizhang/CBD}} to 
implement \CBD, where we set the hyperparameters to the default values.

\section{Precision and Recall of \knnParameter-NN(\NumThreshold) Detectors}  \label{sec:app:knn}

We show the precision and recall of adapted \knnParameter-NN detectors with different options of \NumThreshold, applied 
to select reference data in experiments of purification as a defense on CIFAR-10, purification as a defense on TinyImageNet, 
purification as an attack on CIFAR-100, and purification as an attack on TinyImageNet. These information are listed 
in \tblref{table:poisoning:knn-kappa:cifar10:parameters}, \tblref{table:poisoning:knn-kappa:tinyimagenet:parameters},
\tblref{table:backdoor:knn-kappa:cifar10:parameters}, \tblref{table:backdoor:knn-kappa:tinyimagenet:parameters}, 
\tblref{table:tracing:knn-kappa:cifar100:parameters}, and \tblref{table:tracing:knn-kappa:tinyimagenet:parameters}.

\begin{table} [t!]
\centering
\vspace{0.1in}
\resizebox{0.47\textwidth}{!}{
    \begin{tabular}{@{}lr@{\hspace{1.0em}}rr@{\hspace{1.0em}}rr@{\hspace{1.0em}}rr@{\hspace{1.0em}}rr@{\hspace{1.0em}}r@{}}
    \toprule
         \multicolumn{2}{c}{$\NumThreshold=500$} & \multicolumn{2}{c}{$\NumThreshold=1000$} & \multicolumn{2}{c}{$\NumThreshold=2000$}  & \multicolumn{2}{c}{$\NumThreshold=3000$}  & \multicolumn{2}{c}{$\NumThreshold=4000$} \\
            \multicolumn{1}{c}{\Precision \%} & \multicolumn{1}{c}{\Recall \%} & \multicolumn{1}{c}{\Precision \%} & \multicolumn{1}{c}{\Recall \%} & \multicolumn{1}{c}{\Precision \%} & \multicolumn{1}{c}{\Recall \%}  & \multicolumn{1}{c}{\Precision \%} & \multicolumn{1}{c}{\Recall \%}  & \multicolumn{1}{c}{\Precision \%} & \multicolumn{1}{c}{\Recall \%}  \\
    \midrule
     $98.30$ & $9.92$ & $98.50$ & $19.90$ & $98.72$ & $39.88$ & $98.88$ & $59.92$ & $98.99$ & $79.99$ \\
    \bottomrule
    \end{tabular}
    }
    \caption{Precision and recall of adapted \knnParameter-NN(\NumThreshold) detectors against targeted data poisoning on CIFAR-10.
    }
    \label{table:poisoning:knn-kappa:cifar10:parameters}
\end{table}

\begin{table} [t!]
\centering
\vspace{0.1in}
\resizebox{0.47\textwidth}{!}{
    \begin{tabular}{@{}lr@{\hspace{1.0em}}rr@{\hspace{1.0em}}rr@{\hspace{1.0em}}rr@{\hspace{1.0em}}rr@{\hspace{1.0em}}r@{}}
    \toprule
         \multicolumn{2}{c}{$\NumThreshold=50$} & \multicolumn{2}{c}{$\NumThreshold=100$} & \multicolumn{2}{c}{$\NumThreshold=150$}  & \multicolumn{2}{c}{$\NumThreshold=200$}  & \multicolumn{2}{c}{$\NumThreshold=250$} \\
            \multicolumn{1}{c}{\Precision \%} & \multicolumn{1}{c}{\Recall \%} & \multicolumn{1}{c}{\Precision \%} & \multicolumn{1}{c}{\Recall \%} & \multicolumn{1}{c}{\Precision \%} & \multicolumn{1}{c}{\Recall \%}  & \multicolumn{1}{c}{\Precision \%} & \multicolumn{1}{c}{\Recall \%}  & \multicolumn{1}{c}{\Precision \%} & \multicolumn{1}{c}{\Recall \%}  \\
    \midrule
     $98.20$ & $9.91$ & $98.35$ & $19.87$ & $98.49$ & $29.84$ & $98.60$ & $39.84$ & $98.71$ & $49.85$ \\
    \bottomrule
    \end{tabular}
    }
    \caption{Precision and recall of adapted \knnParameter-NN(\NumThreshold) detectors against targeted data poisoning on TinyImageNet.
    }
    \label{table:poisoning:knn-kappa:tinyimagenet:parameters}
\end{table}


\begin{table} [t!]
    \centering
    \vspace{0.1in}
    \resizebox{0.47\textwidth}{!}{
        \begin{tabular}{@{}lr@{\hspace{1.0em}}rr@{\hspace{1.0em}}rr@{\hspace{1.0em}}rr@{\hspace{1.0em}}rr@{\hspace{1.0em}}r@{}}
        \toprule
            \multicolumn{2}{c}{$\NumThreshold=500$} & \multicolumn{2}{c}{$\NumThreshold=1000$} & \multicolumn{2}{c}{$\NumThreshold=2000$}  & \multicolumn{2}{c}{$\NumThreshold=3000$}  & \multicolumn{2}{c}{$\NumThreshold=4000$} \\
                \multicolumn{1}{c}{\Precision \%} & \multicolumn{1}{c}{\Recall \%} & \multicolumn{1}{c}{\Precision \%} & \multicolumn{1}{c}{\Recall \%} & \multicolumn{1}{c}{\Precision \%} & \multicolumn{1}{c}{\Recall \%}  & \multicolumn{1}{c}{\Precision \%} & \multicolumn{1}{c}{\Recall \%}  & \multicolumn{1}{c}{\Precision \%} & \multicolumn{1}{c}{\Recall \%}  \\
        \midrule
         $99.42$ & $10.04$ & $99.44$ & $20.08$ & $99.44$ & $40.18$ & $99.43$ & $60.26$ & $99.37$ & $80.30$ \\
        \bottomrule
        \end{tabular}
        }
        \caption{Precision and recall of adapted \knnParameter-NN(\NumThreshold) against backdoor attack on CIFAR-10.
        }
        \label{table:backdoor:knn-kappa:cifar10:parameters}
    \end{table}
    
    \begin{table} [t!]
    \centering
    \vspace{0.1in}
    \resizebox{0.47\textwidth}{!}{
        \begin{tabular}{@{}lr@{\hspace{1.0em}}rr@{\hspace{1.0em}}rr@{\hspace{1.0em}}rr@{\hspace{1.0em}}rr@{\hspace{1.0em}}r@{}}
        \toprule
             \multicolumn{2}{c}{$\NumThreshold=50$} & \multicolumn{2}{c}{$\NumThreshold=100$} & \multicolumn{2}{c}{$\NumThreshold=150$}  & \multicolumn{2}{c}{$\NumThreshold=200$}  & \multicolumn{2}{c}{$\NumThreshold=250$}   \\
                \multicolumn{1}{c}{\Precision \%} & \multicolumn{1}{c}{\Recall \%} & \multicolumn{1}{c}{\Precision \%} & \multicolumn{1}{c}{\Recall \%} & \multicolumn{1}{c}{\Precision \%} & \multicolumn{1}{c}{\Recall \%}  & \multicolumn{1}{c}{\Precision \%} & \multicolumn{1}{c}{\Recall \%}  & \multicolumn{1}{c}{\Precision \%} & \multicolumn{1}{c}{\Recall \%}  \\
        \midrule
         $98.73$ & $9.97$ & $98.93$ & $19.98$ & $99.03$ & $30.01$ & $99.10$ & $40.04$ & $99.13$ & $50.06$  \\
        \bottomrule
        \end{tabular}
        }
        \caption{Precision and recall of adapted \knnParameter-NN(\NumThreshold) against backdoor attack on TinyImageNet.
        }
        \label{table:backdoor:knn-kappa:tinyimagenet:parameters}
    \end{table}


\begin{table} [t!]
\centering
\vspace{0.1in}
\resizebox{0.47\textwidth}{!}{
    \begin{tabular}{@{}lr@{\hspace{1.0em}}rr@{\hspace{1.0em}}rr@{\hspace{1.0em}}rr@{\hspace{1.0em}}rr@{\hspace{1.0em}}r@{}}
    \toprule
        \multicolumn{2}{c}{$\NumThreshold=50$} & \multicolumn{2}{c}{$\NumThreshold=100$} & \multicolumn{2}{c}{$\NumThreshold=150$}  & \multicolumn{2}{c}{$\NumThreshold=200$}  & \multicolumn{2}{c}{$\NumThreshold=250$} \\
            \multicolumn{1}{c}{\Precision \%} & \multicolumn{1}{c}{\Recall \%} & \multicolumn{1}{c}{\Precision \%} & \multicolumn{1}{c}{\Recall \%} & \multicolumn{1}{c}{\Precision \%} & \multicolumn{1}{c}{\Recall \%}  & \multicolumn{1}{c}{\Precision \%} & \multicolumn{1}{c}{\Recall \%}  & \multicolumn{1}{c}{\Precision \%} & \multicolumn{1}{c}{\Recall \%}  \\
    \midrule
     $97.75$ & $10.86$ & $97.15$ & $21.58$ & $96.57$ & $32.19$ & $96.03$ & $42.68$ & $95.45$ & $53.03$ \\
    \bottomrule
    \end{tabular}
    }
    \caption{Precision and recall of adapted \knnParameter-NN(\NumThreshold) against data tracing on CIFAR-100.
    }
    \label{table:tracing:knn-kappa:cifar100:parameters}
\end{table}

\begin{table} [t!]
\centering
\vspace{0.1in}
\resizebox{0.47\textwidth}{!}{
    \begin{tabular}{@{}lr@{\hspace{1.0em}}rr@{\hspace{1.0em}}rr@{\hspace{1.0em}}rr@{\hspace{1.0em}}rr@{\hspace{1.0em}}r@{}}
    \toprule
         \multicolumn{2}{c}{$\NumThreshold=50$} & \multicolumn{2}{c}{$\NumThreshold=100$} & \multicolumn{2}{c}{$\NumThreshold=150$}  & \multicolumn{2}{c}{$\NumThreshold=200$}  & \multicolumn{2}{c}{$\NumThreshold=250$}   \\
            \multicolumn{1}{c}{\Precision \%} & \multicolumn{1}{c}{\Recall \%} & \multicolumn{1}{c}{\Precision \%} & \multicolumn{1}{c}{\Recall \%} & \multicolumn{1}{c}{\Precision \%} & \multicolumn{1}{c}{\Recall \%}  & \multicolumn{1}{c}{\Precision \%} & \multicolumn{1}{c}{\Recall \%}  & \multicolumn{1}{c}{\Precision \%} & \multicolumn{1}{c}{\Recall \%}  \\
    \midrule
     $93.70$ & $10.41$ & $93.13$ & $20.69$ & $92.62$ & $30.87$ & $92.20$ & $40.97$ & $91.86$ & $51.03$  \\
    \bottomrule
    \end{tabular}
    }
    \caption{Precision and recall of adapted \knnParameter-NN(\NumThreshold) against data tracing on TinyImageNet.
    }
    \label{table:tracing:knn-kappa:tinyimagenet:parameters}
\end{table}


\section{Additional Experimental Results of Using Purification as
a Defense Against Data Poisoning} \label{sec:app:extra_results:poisoning}

\subsection{CIFAR-10}


\begin{table*} [ht!]
    \centering
    \vspace{0.1in}
    \resizebox{\textwidth}{!}{
        \begin{tabular}{@{}lr@{\hspace{1.5em}}rr@{\hspace{1.5em}}rr@{\hspace{1.5em}}rr@{\hspace{1.5em}}rr@{\hspace{1.5em}}rr@{\hspace{1.5em}}rr@{\hspace{1.5em}}r@{}}
        \toprule
            \multicolumn{1}{c}{\multirow{2}{*}{}} & \multicolumn{1}{c}{} & \multicolumn{2}{c}{$\Precision=0.991$} & \multicolumn{2}{c}{$\Precision=0.993$} & \multicolumn{2}{c}{$\Precision=0.995$}  & \multicolumn{2}{c}{$\Precision=0.998$}  & \multicolumn{2}{c}{$\Precision=1.0$} \\
              & & \multicolumn{1}{c}{\Accuracy \%} & \multicolumn{1}{c}{\AttackSuccessRate} & \multicolumn{1}{c}{\Accuracy \%} & \multicolumn{1}{c}{\AttackSuccessRate} & \multicolumn{1}{c}{\Accuracy \%} & \multicolumn{1}{c}{\AttackSuccessRate}  & \multicolumn{1}{c}{\Accuracy \%} & \multicolumn{1}{c}{\AttackSuccessRate}  & \multicolumn{1}{c}{\Accuracy \%} & \multicolumn{1}{c}{\AttackSuccessRate}  \\
        \midrule
        \multirow{6}{*}{$\Recall=0.025$}& \ReferenceMethod & $31.40(\pm4.83)$ & $1/20$ & $32.29(\pm4.70)$ & $1/20$ & $31.80(\pm4.84)$ & $1/20$ & $32.49(\pm5.70)$ & $2/20$ & $30.82(\pm5.66)$ & $1/20$ \\
        & \RandomNoise & $\mathbf{90.31(\pm0.30)}$ & $2/20$ & $\mathbf{90.22(\pm0.30)}$ & $4/20$ & $\mathbf{90.16(\pm0.38)}$ & $2/20$ & $\mathbf{90.21(\pm0.30)}$ & $2/20$ & $\mathbf{90.34(\pm0.46)}$ & $3/20$ \\
        & \RandomNoiseAlt & $\mathbf{90.19(\pm0.44)}$ & $2/20$ &$\mathbf{90.19(\pm0.44)}$ & $2/20$ & $\mathbf{90.19(\pm0.44)}$ & $2/20$ & $\mathbf{90.19(\pm0.44)}$ & $2/20$ & $\mathbf{90.19(\pm0.44)}$ & $2/20$ \\
        & \FriendlyNoise & $\mathbf{90.16(\pm0.18)}$ & $3/20$ & $\mathbf{90.46(\pm0.25)}$ & $3/20$ & $\mathbf{90.32(\pm0.36)}$  & $3/20$ & $\mathbf{90.19(\pm0.38)}$ & $2/20$ & $\mathbf{90.28(\pm0.31)}$ & $2/20$\\
        & \FriendlyNoiseAlt & $\mathbf{90.00(\pm0.42)}$ & $1/20$ & $\mathbf{90.00(\pm0.42)}$ & $\mathbf{1/20}$ & $\mathbf{90.00(\pm0.42)}$ & $\mathbf{1/20}$ & $\mathbf{90.00(\pm0.42)}$ & $\mathbf{1/20}$ & $\mathbf{90.00(\pm0.42)}$ & $1/20$ \\
        & \sysName & $\mathbf{90.63(\pm0.45)}$ & $\mathbf{0/20}$ & $\mathbf{90.54(\pm0.61)}$ & $\mathbf{1/20}$ & $\mathbf{90.80(\pm0.49)}$  & $\mathbf{1/20}$ & $\mathbf{90.73(\pm0.54)}$ & $\mathbf{1/20}$ & $\mathbf{90.59(\pm0.48)}$ & $\mathbf{0/20}$\\ \midrule
        
        \multirow{6}{*}{$\Recall=0.050$}& \ReferenceMethod &  $42.04(\pm4.12)$ & $0/20$ & $44.70(\pm4.01)$ & $0/20$ & $44.81(\pm4.09)$ & $1/20$ & $42.51(\pm3.20)$ & $1/20$ & $43.73(\pm3.75)$ & $0/20$ \\
        & \RandomNoise & $90.27(\pm0.37)$ & $3/20$ & $90.27(\pm0.29)$ & $0/20$ & $90.30(\pm0.48)$ & $2/20$ & $\mathbf{90.32(\pm0.34)}$ & $2/20$ & $\mathbf{90.45(\pm0.37)}$ & $2/20$ \\
       & \RandomNoiseAlt & $90.19(\pm0.44)$ & $2/20$ &$90.19(\pm0.44)$ & $2/20$ & $90.19(\pm0.44)$ & $2/20$ & $\mathbf{90.19(\pm0.44)}$ & $2/20$ & $90.19(\pm0.44)$ & $2/20$ \\
        & \FriendlyNoise & $\mathbf{90.29(\pm0.40)}$ & $3/20$ & $\mathbf{90.45(\pm0.24)}$ & $4/20$ & $\mathbf{90.41(\pm0.32)}$  & $3/20$ & $\mathbf{90.36(\pm0.33)}$ & $2/20$ & $\mathbf{90.36(\pm0.37)}$  & $2/20$\\
        & \FriendlyNoiseAlt & ${90.00(\pm0.42)}$ & ${1/20}$ & $90.00(\pm0.42)$ & $1/20$ & ${90.00(\pm0.42)}$ & ${1/20}$ & $\mathbf{90.00(\pm0.42)}$ & $\mathbf{1/20}$ & $90.00(\pm0.42)$ & $1/20$ \\
        & \sysName & $\mathbf{91.28(\pm0.47)}$ & $\mathbf{1/20}$ & $\mathbf{91.32(\pm0.54)}$ & $\mathbf{0/20}$ & $\mathbf{91.36(\pm0.42)}$  & $\mathbf{1/20}$ & $\mathbf{90.97(\pm0.61)}$ & $\mathbf{1/20}$ & $\mathbf{91.34(\pm0.37)}$  & $\mathbf{0/20}$\\  \midrule
    
        \multirow{6}{*}{$\Recall=0.075$}& \ReferenceMethod & $50.45(\pm4.11)$ & $0/20$ & $48.39(\pm4.31)$ & $1/20$ & $49.08(\pm4.01)$ & $0/20$ & $52.61(\pm4.66)$ & $0/20$ & $50.77(\pm3.54)$ & $0/20$ \\
       & \RandomNoise & $\mathbf{90.38(\pm0.38)}$ & $4/20$ & $\mathbf{90.48(\pm0.32)}$ & $4/20$ & $\mathbf{90.40(\pm0.38)}$ & $3/20$ & $\mathbf{90.53(\pm0.32)}$ & $2/20$ & $\mathbf{90.56(\pm0.37)}$ & $4/20$ \\
        & \RandomNoiseAlt & $\mathbf{90.19(\pm0.44)}$ & $2/20$ &$\mathbf{90.19(\pm0.44)}$ & $2/20$ & $90.19(\pm0.44)$ & $2/20$ & $\mathbf{90.19(\pm0.44)}$ & $2/20$ & $90.19(\pm0.44)$ & $2/20$ \\
        & \FriendlyNoise & $\mathbf{90.50(\pm0.40)}$ & $4/20$ & $\mathbf{90.49(\pm0.37)}$ & $3/20$ & $\mathbf{90.38(\pm0.37)}$ & $4/20$ & $\mathbf{90.33(\pm0.34)}$ & $\mathbf{1/20}$ & $90.36(\pm0.31)$ & $4/20$ \\
        & \FriendlyNoiseAlt & $90.00(\pm0.42)$ & $1/20$ & ${90.00(\pm0.42)}$ & ${1/20}$ & ${90.00(\pm0.42)}$ & ${1/20}$ & ${90.00(\pm0.42)}$ & ${1/20}$ & $90.00(\pm0.42)$ & $1/20$ \\
        & \sysName & $\mathbf{91.14(\pm0.44)}$ & $\mathbf{0/20}$ & $\mathbf{91.13(\pm0.35)}$ & $\mathbf{1/20}$ & $\mathbf{91.21(\pm0.51)}$  & $\mathbf{1/20}$ & $\mathbf{91.16(\pm0.53)}$ & $\mathbf{1/20}$ & $\mathbf{91.38(\pm0.45)}$ & $\mathbf{0/20}$ \\
    
        \bottomrule
        \end{tabular}
        }
        \caption{Additional experimental results of purification against \emph{targeted attack} on CIFAR-10 where simulated detectors parameterized by precision $\Precision$ and recall $\Recall$ were applied. \GassianVariant and \PerturbationMagnitude  in baselines were
          searched such that their \Accuracy are as close as possible to \Accuracy of \sysName. We bold the numbers following the same rule as \tblref{table:poisoning:human:cifar10}.
        }
        \label{table:poisoning:human:cifar10:extra}
    \end{table*}

In the experiments of implementing \sysName as a defense against data poisoning 
attack on CIFAR-10, we explored using a simulated detector with a smaller recall 
$\Recall < 0.1$ to select reference data. Such simulated detector still can detect 
an enough amount of reference data per class. For example, a detector with 
$\Recall = 0.025$ and $\Precision = 1.0$ can detect $123$ samples per class, 
which can provide enough information to purify untrusted data. The results 
in \tblref{table:poisoning:human:cifar10:extra} show that \sysName in these 
settings outperformed the other baselines. As such, \sysName is still effective 
even when there was a small amount of reference data available.

\subsection{TinyImageNet}

\begin{table*} [ht!]
    \centering
    \vspace{0.1in}
        
        {\resizebox{\textwidth}{!}{\begin{tabular}{@{}lr@{\hspace{1.5em}}rr@{\hspace{1.5em}}rr@{\hspace{1.5em}}rr@{\hspace{1.5em}}rr@{\hspace{1.5em}}rr@{\hspace{1.5em}}rr@{\hspace{1.5em}}r@{}}
        \toprule
            \multicolumn{1}{c}{\multirow{2}{*}{}} & \multicolumn{1}{c}{} & \multicolumn{2}{c}{$\Precision=0.991$} & \multicolumn{2}{c}{$\Precision=0.993$} & \multicolumn{2}{c}{$\Precision=0.995$}  & \multicolumn{2}{c}{$\Precision=0.998$}  & \multicolumn{2}{c}{$\Precision=1.0$} \\
              & & \multicolumn{1}{c}{\Accuracy \%} & \multicolumn{1}{c}{\AttackSuccessRate} & \multicolumn{1}{c}{\Accuracy \%} & \multicolumn{1}{c}{\AttackSuccessRate} & \multicolumn{1}{c}{\Accuracy \%} & \multicolumn{1}{c}{\AttackSuccessRate}  & \multicolumn{1}{c}{\Accuracy \%} & \multicolumn{1}{c}{\AttackSuccessRate}  & \multicolumn{1}{c}{\Accuracy \%} & \multicolumn{1}{c}{\AttackSuccessRate}  \\
        \midrule
    
        \multirow{4}{*}{$\Recall=0.1$}& \ReferenceMethod &$22.89(\pm1.61)$ & $0/20$ & $22.30(\pm2.48)$ & $0/20$ & $23.75(\pm1.44)$ & $0/20$ & $22.98(\pm1.65)$ & $1/20$ & $20.36(\pm2.28)$ & $0/20$ \\ 
       & \RandomNoise &$\mathbf{59.05(\pm0.46)}$ & $16/20$ & $\mathbf{59.32(\pm0.85)}$ & $15/20$ & $\mathbf{59.35(\pm0.91)}$ & $16/20$ & $\mathbf{59.28(\pm0.77)}$ & $15/20$ & $\mathbf{59.00(\pm0.87)}$ & $16/20$ \\  
        & \FriendlyNoise &$\mathbf{59.12(\pm0.73)}$ & $15/20$ & $\mathbf{59.49(\pm0.88)}$ & $17/20$ & $\mathbf{58.82(\pm0.99)}$ & $17/20$ & $\mathbf{59.47(\pm0.67)}$ & $17/20$ & $\mathbf{59.17(\pm0.90)}$ & $15/20$ \\
        & \sysName &$\mathbf{59.87(\pm0.81)}$ & $\mathbf{3/20}$ & $\mathbf{59.30(\pm0.74)}$ & $\mathbf{3/20}$ & $\mathbf{59.65(\pm0.86)}$ & $\mathbf{1/20}$ & $\mathbf{59.67(\pm0.80)}$ & $\mathbf{1/20}$ & $\mathbf{59.31(\pm1.04)}$ & $\mathbf{1/20}$ \\ 
        \midrule
    
        \multirow{4}{*}{$\Recall=0.2$}& \ReferenceMethod & $27.53(\pm2.04)$ & $0/20$ & $27.15(\pm1.84)$ & $0/20$ & $27.71(\pm1.91)$ & $0/20$ & $27.24(\pm2.40)$ & $0/20$ & $33.74(\pm1.18)$ & $0/20$ \\ 
        & \RandomNoise &$\mathbf{59.61(\pm0.88)}$ & $16/20$ & $\mathbf{59.59(\pm0.73)}$ & $15/20$ & $\mathbf{60.39(\pm1.13)}$ & $17/20$ & $\mathbf{60.72(\pm0.80)}$ & $16/20$ & $\mathbf{60.45(\pm0.82)}$ & $18/20$\\  
        & \FriendlyNoise &$\mathbf{59.86(\pm0.86)}$ & $15/20$ & $\mathbf{59.90(\pm0.66)}$ & $16/20$ & $\mathbf{59.83(\pm1.09)}$ & $17/20$ & $\mathbf{59.89(\pm0.80)}$ & $18/20$ & $\mathbf{59.53(\pm0.74)}$ & $17/20$ \\
        & \sysName &$\mathbf{59.71(\pm0.82)}$ & $\mathbf{8/20}$ & $\mathbf{59.89(\pm0.65)}$ & $\mathbf{5/20}$ & $\mathbf{60.45(\pm0.88)}$ & $\mathbf{6/20}$ & $\mathbf{60.04(\pm0.63)}$ & $\mathbf{0/20}$ & $\mathbf{60.38(\pm0.71)}$ & $\mathbf{2/20}$ \\ 
        \midrule
    
        \multirow{4}{*}{$\Recall=0.3$}& \ReferenceMethod & $41.36(\pm1.16)$ & $2/20$ & $41.39(\pm1.17)$ & $2/20$ & $40.87(\pm1.31)$ & $1/20$ & $41.58(\pm1.68)$ & $0/20$ & $41.14(\pm1.22)$ & $0/20$ \\
        & \RandomNoise &$\mathbf{60.42(\pm0.69)}$ & $17/20$ & $\mathbf{60.50(\pm0.97)}$ & $\mathbf{16/20}$ & $\mathbf{60.71(\pm0.83)}$ & $16/20$ & $\mathbf{60.37(\pm0.95)}$ & $17/20$ & $\mathbf{60.58(\pm0.61)}$ & $17/20$ \\ 
        & \FriendlyNoise &$\mathbf{60.11(\pm0.79)}$ & $16/20$ & $\mathbf{59.91(\pm1.04)}$ & $\mathbf{16/20}$ & $\mathbf{60.33(\pm0.83)}$ & $17/20$ & $\mathbf{60.06(\pm0.69)}$ & $17/20$ & $\mathbf{60.19(\pm1.03)}$ & $17/20$ \\
        & \sysName &$\mathbf{60.53(\pm0.62)}$ & $\mathbf{12/20}$ & $\mathbf{60.55(\pm0.76)}$ & $\mathbf{16/20}$ & $\mathbf{60.15(\pm0.99)}$ & $\mathbf{9/20}$ & $\mathbf{60.17(\pm0.75)}$ & $\mathbf{4/20}$ & $\mathbf{60.28(\pm0.95)}$ & $\mathbf{0/20}$ \\ 
        \bottomrule
        \end{tabular}
        }}
    
        \caption{Experimental results of purification against \emph{targeted attack} on TinyImageNet where simulated detectors parameterized by precision
          $\Precision$ and recall $\Recall$ were applied. We bold the numbers following the same
          rule as \tblref{table:poisoning:human:cifar10}.}
        \label{table:poisoning:human:tiny}
    \end{table*}
    
    \begin{table*} [ht!]
    \centering
    \vspace{0.1in}
    
        {\resizebox{\textwidth}{!}{
        \begin{tabular}{@{}lr@{\hspace{0.3em}}rr@{\hspace{0.3em}}rr@{\hspace{0.3em}}rr@{\hspace{0.3em}}rr@{\hspace{0.3em}}rr@{\hspace{0.3em}}r@{}}
        \toprule
            \multicolumn{1}{c}{\multirow{2}{*}{}} & \multicolumn{2}{c}{$\NumThreshold=50$} & \multicolumn{2}{c}{$\NumThreshold=100$} & \multicolumn{2}{c}{$\NumThreshold=150$}  & \multicolumn{2}{c}{$\NumThreshold=200$}  & \multicolumn{2}{c}{$\NumThreshold=250$} \\
               & \multicolumn{1}{c}{\Accuracy \%} & \multicolumn{1}{c}{$\quad\AttackSuccessRate\quad$} & \multicolumn{1}{c}{\Accuracy \%} & \multicolumn{1}{c}{$\quad\AttackSuccessRate\quad$} & \multicolumn{1}{c}{\Accuracy \%} & \multicolumn{1}{c}{$\quad\AttackSuccessRate\quad$}  & \multicolumn{1}{c}{\Accuracy \%} & \multicolumn{1}{c}{$\quad\AttackSuccessRate\quad$}  & \multicolumn{1}{c}{\Accuracy \%} & \multicolumn{1}{c}{$\quad\AttackSuccessRate\quad$}  \\
        \midrule
        \ReferenceMethod & $28.25(\pm1.29)$ & $0/20$ & $33.06(\pm1.57)$ & $0/20$ & $43.78(\pm0.88)$ & $0/20$ & $47.87(\pm1.30)$ & $1/20$ & $52.20(\pm0.78)$ & $0/20$ \\
        \RandomNoise &  $\mathbf{54.76(\pm0.68)}$ & $8/20$ & $\mathbf{56.48(\pm0.79)}$ & $9/20$ & $\mathbf{56.77(\pm0.64)}$ & $11/20$ & $\mathbf{57.22(\pm0.92)}$ & $9/20$ & $\mathbf{58.43(\pm0.86)}$ & $14/20$ \\
        \FriendlyNoise & $\mathbf{54.91(\pm0.63)}$ & $7/20$ & $\mathbf{56.55(\pm0.82)}$ & $10/20$ & $\mathbf{56.60(\pm0.57)}$ & $9/20$ & $\mathbf{57.40(\pm0.94)}$ & $11/20$ & $\mathbf{58.69(\pm0.82)}$ & $14/20$ \\
        \sysName & $\mathbf{54.74(\pm0.78)}$ & $\mathbf{1/20}$ & $\mathbf{56.03(\pm0.76)}$ & $\mathbf{1/20}$ & $\mathbf{56.91(\pm0.80)}$ & $\mathbf{0/20}$ & $\mathbf{57.90(\pm0.74)}$ & $\mathbf{1/20}$ & $\mathbf{58.32(\pm0.66)}$ & $\mathbf{3/20}$ \\
        \bottomrule
        \end{tabular}
        }}
    
        \caption{Experimental results of purification against \emph{targeted attack} on TinyImageNet where \knnParameter-NN(\NumThreshold) detectors were
          applied. The precision and recall of such detectors are shown in
          \tblref{table:poisoning:knn-kappa:tinyimagenet:parameters} in
          \appref{sec:app:knn}. We bold the numbers following the same
          rule as \tblref{table:poisoning:human:cifar10}.}
        \label{table:poisoning:knn-kappa:tiny}
    \end{table*}

    
    \begin{table*} [ht!]
      \centering
      \vspace{0.1in}
          
          {\resizebox{\textwidth}{!}{\begin{tabular}{@{}lr@{\hspace{1.5em}}rr@{\hspace{1.5em}}rr@{\hspace{1.5em}}rr@{\hspace{1.5em}}rr@{\hspace{1.5em}}rr@{\hspace{1.5em}}rr@{\hspace{1.5em}}r@{}}
          \toprule
              \multicolumn{1}{c}{\multirow{2}{*}{}} & \multicolumn{1}{c}{} & \multicolumn{2}{c}{$\Precision=0.991$} & \multicolumn{2}{c}{$\Precision=0.993$} & \multicolumn{2}{c}{$\Precision=0.995$}  & \multicolumn{2}{c}{$\Precision=0.998$}  & \multicolumn{2}{c}{$\Precision=1.0$} \\
                & & \multicolumn{1}{c}{\Accuracy \%} & \multicolumn{1}{c}{\AttackSuccessRate \%} & \multicolumn{1}{c}{\Accuracy \%} & \multicolumn{1}{c}{\AttackSuccessRate \%} & \multicolumn{1}{c}{\Accuracy \%} & \multicolumn{1}{c}{\AttackSuccessRate \%}  & \multicolumn{1}{c}{\Accuracy \%} & \multicolumn{1}{c}{\AttackSuccessRate \%}  & \multicolumn{1}{c}{\Accuracy \%} & \multicolumn{1}{c}{\AttackSuccessRate \%}  \\
          \midrule
      
          \multirow{4}{*}{$\Recall=0.1$}& \ReferenceMethod & $20.59(\pm2.05)$ & $1.59$ & $21.54(\pm1.47)$ & $0.99$ & $20.74(\pm1.73)$ & $0.69$ & $21.18(\pm1.92)$ & $0.79$ & $21.51(\pm1.08)$ & $0.69$\\
          & \RandomNoise & $\mathbf{60.00(\pm0.73)}$ & $68.89$ & $\mathbf{59.54(\pm0.57)}$ & $66.79$ & $\mathbf{59.51(\pm0.66)}$ & $66.69$ & $\mathbf{59.92(\pm0.68)}$ & $67.09$ & $\mathbf{59.40(\pm0.90)}$ & $65.69$\\ 
          & \FriendlyNoise & $\mathbf{59.49(\pm0.69)}$ & $67.69$ & $\mathbf{59.51(\pm0.71)}$ & $67.69$ & $\mathbf{59.40(\pm0.63)}$ & $68.59$ & $\mathbf{59.68(\pm0.87)}$ & $69.29$ & $\mathbf{59.64(\pm0.88)}$ & $67.89$\\ 
          & \sysName &$\mathbf{59.42(\pm0.77)}$ & $\mathbf{12.19}$ & $\mathbf{59.40(\pm0.88)}$ & $\mathbf{9.29}$ & $\mathbf{59.23(\pm0.88)}$ & $\mathbf{7.39}$ & $\mathbf{59.60(\pm1.09)}$ & $\mathbf{4.69}$ & $\mathbf{59.24(\pm0.99)}$ & $\mathbf{3.89}$ \\ 
          \midrule
      
          \multirow{4}{*}{$\Recall=0.2$} & \ReferenceMethod  & $33.60(\pm1.76)$ & $0.79$ & $34.09(\pm0.92)$ & $1.69$ & $33.43(\pm1.74)$ & $1.89$ & $33.82(\pm1.26)$ & $0.49$ & $34.16(\pm1.16)$ & $0.59$\\
          & \RandomNoise & $\mathbf{59.60(\pm0.90)}$ & $71.79$ & $\mathbf{59.70(\pm0.81)}$ & $71.59$ & $\mathbf{59.66(\pm0.75)}$ & $68.89$ & $\mathbf{59.72(\pm0.73)}$ & $71.89$ & $\mathbf{59.51(\pm0.74)}$ & $72.09$\\
          & \FriendlyNoise  & $\mathbf{60.28(\pm0.87)}$ & $72.29$ & $\mathbf{60.55(\pm1.01)}$ & $70.59$ & $\mathbf{60.66(\pm0.90)}$ & $70.79$ & $\mathbf{60.68(\pm0.93)}$ & $70.29$ & $\mathbf{60.37(\pm0.79)}$ & $69.89$\\
          & \sysName & $\mathbf{60.00(\pm0.86)}$ & $\mathbf{22.09}$ & $\mathbf{59.68(\pm1.11)}$ & $\mathbf{18.59}$ & $\mathbf{59.91(\pm0.74)}$ & $\mathbf{13.49}$ & $\mathbf{60.31(\pm0.73)}$ & $\mathbf{5.89}$ & $\mathbf{60.00(\pm0.75)}$ & $\mathbf{3.39}$ \\ 
          \midrule
      
          \multirow{4}{*}{$\Recall=0.3$} & \ReferenceMethod  & $41.17(\pm1.11)$ & $6.39$ & $40.85(\pm1.37)$ & $3.19$ & $41.39(\pm1.28)$ & $1.69$ & $41.27(\pm1.37)$ & $0.89$ & $41.47(\pm1.11)$ & $0.79$\\
          & \RandomNoise & $\mathbf{59.76(\pm0.79)}$ & $71.49$ & $59.25(\pm0.74)$ & $73.79$ & $59.18(\pm0.86)$ & $70.59$ & $\mathbf{59.48(\pm0.76)}$ & $73.39$ & $\mathbf{59.79(\pm0.64)}$ & $74.59$\\
          & \FriendlyNoise & $\mathbf{60.27(\pm1.03)}$ & $74.79$ & $\mathbf{60.68(\pm0.92)}$ & $74.29$ & $\mathbf{60.51(\pm0.62)}$ & $74.19$ & $\mathbf{60.32(\pm0.89)}$ & $70.69$ & $\mathbf{60.20(\pm0.83)}$ & $71.19$\\
          & \sysName & $\mathbf{60.39(\pm0.76)}$ & $\mathbf{26.39}$ & $\mathbf{60.41(\pm0.91)}$ & $\mathbf{23.09}$ & $\mathbf{60.50(\pm0.86)}$ & $\mathbf{15.79}$ & $\mathbf{60.44(\pm0.81)}$ & $\mathbf{6.59}$ & $\mathbf{60.56(\pm0.77)}$ & $\mathbf{3.49}$ \\
          \bottomrule
          \end{tabular}
          }}
      
          \caption{Experimental results of purification against \emph{backdoor attack} on TinyImageNet where simulated detectors parameterized by precision
            $\Precision$ and recall $\Recall$ were applied. We bold the numbers following the same
            rule as \tblref{table:poisoning:human:cifar10}.}
          \label{table:backdoor:human:tiny}
      \end{table*}

      \begin{table*} [ht!]
          \centering
          \vspace{0.1in}
          
              {\resizebox{\textwidth}{!}{
              \begin{tabular}{@{}lr@{\hspace{0.3em}}rr@{\hspace{0.3em}}rr@{\hspace{0.3em}}rr@{\hspace{0.3em}}rr@{\hspace{0.3em}}rr@{\hspace{0.3em}}r@{}}
              \toprule
                  \multicolumn{1}{c}{\multirow{2}{*}{}} & \multicolumn{2}{c}{$\NumThreshold=50$} & \multicolumn{2}{c}{$\NumThreshold=100$} & \multicolumn{2}{c}{$\NumThreshold=150$}  & \multicolumn{2}{c}{$\NumThreshold=200$}  & \multicolumn{2}{c}{$\NumThreshold=250$} \\
                     & \multicolumn{1}{c}{\Accuracy \%} & \multicolumn{1}{c}{\AttackSuccessRate \%} & \multicolumn{1}{c}{\Accuracy \%} & \multicolumn{1}{c}{\AttackSuccessRate \%} & \multicolumn{1}{c}{\Accuracy \%} & \multicolumn{1}{c}{\AttackSuccessRate \%}  & \multicolumn{1}{c}{\Accuracy \%} & \multicolumn{1}{c}{\AttackSuccessRate \%}  & \multicolumn{1}{c}{\Accuracy \%} & \multicolumn{1}{c}{\AttackSuccessRate \%}  \\
              \midrule
              \ReferenceMethod & $28.53(\pm1.27)$ & $0.39$ & $32.57(\pm1.39)$ & $0.39$ & $44.57(\pm0.70)$ & $1.49$ & $48.21(\pm0.71)$ & $2.09$ & $52.06(\pm0.83)$ & $6.69$\\
              \RandomNoise & $\mathbf{55.34(\pm0.85)}$ & $29.29$ & $\mathbf{56.02(\pm0.42)}$ & $35.29$ & $\mathbf{57.37(\pm0.55)}$ & $44.59$ & $\mathbf{57.70(\pm0.60)}$ & $47.89$ & $\mathbf{58.41(\pm0.67)}$ & $52.99$ \\
              \FriendlyNoise & $\mathbf{55.43(\pm1.00)}$ & $34.79$ & $\mathbf{56.66(\pm0.62)}$ & $37.29$ & $\mathbf{57.07(\pm0.80)}$ & $46.69$ & $\mathbf{57.76(\pm0.72)}$ & $55.49$ & $\mathbf{58.78(\pm0.77)}$ & $55.89$ \\
              \sysName & $\mathbf{55.09(\pm0.68)}$ & $\mathbf{2.49}$ & $\mathbf{56.18(\pm0.89)}$ & $\mathbf{3.99}$ & $\mathbf{57.17(\pm0.81)}$ & $\mathbf{6.29}$ & $\mathbf{57.80(\pm0.87)}$ & $\mathbf{5.79}$ & $\mathbf{58.82(\pm0.64)}$ & $\mathbf{7.19}$ \\
              \bottomrule
              \end{tabular}
              }}
          
              \caption{Experimental results of purification against \emph{backdoor attack} on TinyImageNet where \knnParameter-NN(\NumThreshold) detectors were
                applied. The precision and recall of such detectors are shown in
                \tblref{table:backdoor:knn-kappa:tinyimagenet:parameters} in
                \appref{sec:app:knn}. We bold the numbers following the same
                rule as \tblref{table:poisoning:human:cifar10}.}
              \label{table:backdoor:knn-kappa:tiny}
          \end{table*}

The experimental results of implementing \sysName as a defense against data poisoning 
attack on TinyImageNet are shown in \tblref{table:poisoning:human:tiny}, 
\tblref{table:poisoning:knn-kappa:tiny}, \tblref{table:backdoor:human:tiny},
and \tblref{table:backdoor:knn-kappa:tiny}.

\tblref{table:poisoning:human:tiny} and \tblref{table:backdoor:human:tiny} 
show the results of \sysName and
purification baselines (with searched $\GassianVariant$ or
$\PerturbationMagnitude$) against \emph{targeted attack} and \emph{backdoor attack} 
on TinyImageNet when simulated detectors
with different precisions and recalls were used. First, \sysName
achieved the lowest \AttackSuccessRate when achieving comparable
\Accuracy with the baselines in most cases. In other words, when
achieving comparable utility for the purified training data, the
models learned based on the training data purified by \sysName were
less likely to predict the target inputs or inputs with the attacker-chosen trigger
as target labels. Second, an
overall trend is that \sysName was better at eliminating the hidden
properties from the learned models when $\Precision$ was larger or
$\Recall$ was smaller. 

\tblref{table:poisoning:knn-kappa:tiny} and \tblref{table:backdoor:knn-kappa:tiny} show the results when the
\knnParameter-NN($\NumThreshold$) detector with different \NumThreshold was used to
detect the reference data. \sysName outperforms the baselines 
against both targeted attacks and backdoor attacks (i.e., \sysName achieved lower \AttackSuccessRate when
achieving comparable \Accuracy) in all options of \NumThreshold.

These results on TinyImageNet demonstrate the effectiveness 
and advantages of our proposed method against targeted attacks and backdoor attacks.

\section{Additional Experimental Results of Using Purification as 
an Attack on Data Tracing} \label{sec:app:extra_results:tracing}

\subsection{TinyImageNet}

The experimental results of implementing \sysName as an attack on data
tracing on TinyImageNet are shown in \tblref{table:tracing:human:tiny} and
\tblref{table:tracing:knn-kappa:tiny}.

When considering a simulated detector, \sysName outperforms the baselines in most cases. Specifically, when
achieving comparable \Accuracy, \sysName achieves lower frequency of
detection (i.e., $\PValue < 0.05$) in most cases, shown 
in \tblref{table:tracing:human:tiny}. In other words, when
preserving the data utility to the same extent, \sysName is better at
eliminating the hidden properties from the models learned using the
purified training data.  \sysName also achieves lower frequency of
detection (i.e., more likely to eliminate the hidden properties) when
\Recall of the simulated detector is smaller, especially when
\Precision is small.

\tblref{table:tracing:knn-kappa:tiny} shows the results when the
reference data is selected using the adapted \knnParameter-NN detector
with different $\NumThreshold$.  \sysName also outperforms the
baselines when such a detector is used.

          \begin{table*} [ht!]
            \centering
            \vspace{0.1in}
           
                { \resizebox{\textwidth}{!}{
                \begin{tabular}{@{}lr@{\hspace{1.5em}}rr@{\hspace{1.5em}}rr@{\hspace{1.5em}}rr@{\hspace{1.5em}}rr@{\hspace{1.5em}}rr@{\hspace{1.5em}}rr@{\hspace{1.5em}}r@{}}
                \toprule
                    \multicolumn{1}{c}{\multirow{2}{*}{}} & \multicolumn{1}{c}{} & \multicolumn{2}{c}{$\Precision=0.92$} & \multicolumn{2}{c}{$\Precision=0.94$} & \multicolumn{2}{c}{$\Precision=0.96$}  & \multicolumn{2}{c}{$\Precision=0.98$}  & \multicolumn{2}{c}{$\Precision=1.0$} \\
                      & & \multicolumn{1}{c}{\Accuracy \%} & \multicolumn{1}{c}{$\PValue < 0.05$} & \multicolumn{1}{c}{\Accuracy \%} & \multicolumn{1}{c}{$\PValue < 0.05$} & \multicolumn{1}{c}{\Accuracy \%} & \multicolumn{1}{c}{$\PValue < 0.05$}  & \multicolumn{1}{c}{\Accuracy \%} & \multicolumn{1}{c}{$\PValue < 0.05$}  & \multicolumn{1}{c}{\Accuracy \%} & \multicolumn{1}{c}{$\PValue < 0.05$}  \\
                \midrule
                
                \multirow{4}{*}{$\Recall=0.1$}& \ReferenceMethod &$22.82(\pm1.28)$ & $1/20$ & $22.41(\pm1.30)$ & $2/20$ & $22.92(\pm1.08)$ & $2/20$ & $22.33(\pm0.89)$ & $1/20$ & $21.73(\pm1.26)$ & $0/20$  \\  
                & \RandomNoise &$\mathbf{56.54(\pm0.92)}$ & $11/20$ & $\mathbf{56.90(\pm0.55)}$ & $\mathbf{7/20}$ & $\mathbf{56.44(\pm0.94)}$ & $8/20$ & $\mathbf{56.59(\pm0.50)}$ & $9/20$ & $\mathbf{56.62(\pm0.72)}$ & $\mathbf{8/20}$  \\
                & \FriendlyNoise &$\mathbf{56.80(\pm0.73)}$ & $\mathbf{7/20}$ & $\mathbf{56.28(\pm0.74)}$ & $\mathbf{7/20}$ & $\mathbf{56.96(\pm0.81)}$ & $9/20$ & $\mathbf{56.66(\pm0.62)}$ & $\mathbf{7/20}$ & $\mathbf{56.88(\pm0.80)}$ & $10/20$  \\
                & \sysName &$\mathbf{56.62(\pm0.81)}$ & $\mathbf{7/20}$ & $\mathbf{56.84(\pm0.78)}$ & $8/20$ & $\mathbf{56.80(\pm0.69)}$ & $\mathbf{7/20}$ & $\mathbf{56.58(\pm0.88)}$ & $8/20$ & $\mathbf{56.39(\pm0.91)}$ & $\mathbf{8/20}$  \\\midrule
            
                \multirow{4}{*}{$\Recall=0.2$}& \ReferenceMethod & $33.89(\pm1.17)$ & $3/20$ & $30.74(\pm2.09)$ & $1/20$ & $32.55(\pm1.58)$ & $0/20$ & $31.64(\pm1.57)$ & $1/20$ & $31.12(\pm1.33)$ & $0/20$ \\ 
                & \RandomNoise &$\mathbf{57.42(\pm0.71)}$ & $10/20$ & $\mathbf{57.47(\pm0.63)}$ & $12/20$ & $\mathbf{57.93(\pm0.71)}$ & $10/20$ & $\mathbf{57.21(\pm0.88)}$ & $10/20$ & $\mathbf{57.54(\pm0.88)}$ & $12/20$  \\
                & \FriendlyNoise &$\mathbf{57.44(\pm0.68)}$ & $11/20$ & $\mathbf{57.28(\pm0.66)}$ & $13/20$ & $\mathbf{57.45(\pm0.80)}$ & $11/20$ & $\mathbf{57.62(\pm0.64)}$ & $\mathbf{9/20}$ & $\mathbf{57.62(\pm0.78)}$ & $10/20$  \\
                & \sysName &$\mathbf{57.34(\pm0.81)}$ & $\mathbf{8/20}$ & $\mathbf{57.29(\pm0.82)}$ & $\mathbf{8/20}$ & $\mathbf{57.39(\pm0.74)}$ & $\mathbf{6/20}$ & $\mathbf{57.31(\pm0.85)}$ & $\mathbf{9/20}$ & $\mathbf{57.59(\pm0.91)}$ & $\mathbf{8/20}$  \\ \midrule
            
                \multirow{4}{*}{$\Recall=0.3$}& \ReferenceMethod & $28.39(\pm2.52)$ & $1/20$ & $40.89(\pm1.13)$ & $3/20$ & $40.30(\pm1.27)$ & $3/20$ & $40.54(\pm1.02)$ & $2/20$ & $39.50(\pm1.55)$ & $3/20$ \\ 
                & \RandomNoise &$\mathbf{58.39(\pm0.74)}$ & $13/20$ & $\mathbf{57.78(\pm0.64)}$ & $11/20$ & $\mathbf{57.48(\pm0.76)}$ & $12/20$ & $\mathbf{57.51(\pm0.62)}$ & $\mathbf{11/20}$ & $\mathbf{57.67(\pm0.91)}$ & $11/20$  \\
                & \FriendlyNoise &$\mathbf{58.34(\pm0.82)}$ & $12/20$ & $\mathbf{57.73(\pm0.66)}$ & $12/20$ & $\mathbf{57.97(\pm0.76)}$ & $\mathbf{10/20}$ & $\mathbf{57.73(\pm0.55)}$ & $\mathbf{11/20}$ & $\mathbf{57.81(\pm0.70)}$ & $14/20$  \\
                & \sysName &$\mathbf{58.05(\pm0.85)}$ & $\mathbf{10/20}$ & $\mathbf{57.82(\pm0.92)}$ & $\mathbf{8/20}$ & $\mathbf{57.54(\pm0.78)}$ & $11/20$ & $\mathbf{57.96(\pm0.53)}$ & $\mathbf{11/20}$ & $\mathbf{57.80(\pm0.87)}$ & $\mathbf{7/20}$  \\
                \bottomrule
                \end{tabular}
                }}
            
                \caption{Experimental results of purification against \emph{data tracing}
                  on TinyImageNet where
                  simulated detectors parameterized by precision $\Precision$ and
                  recall $\Recall$ were applied. \GassianVariant and
                  \PerturbationMagnitude in baselines were searched such that
                  their \Accuracy are as close as possible to \Accuracy of
                  \sysName. We bold the numbers following the same rule as
                  \tblref{table:poisoning:human:cifar10}.
                }
                \label{table:tracing:human:tiny}
            \end{table*}

            \begin{table*} [ht!]
            \centering
            \vspace{0.1in}
            
                {\resizebox{\textwidth}{!}{
                \begin{tabular}{@{}lr@{\hspace{0.3em}}rr@{\hspace{0.3em}}rr@{\hspace{0.3em}}rr@{\hspace{0.3em}}rr@{\hspace{0.3em}}rr@{\hspace{0.3em}}r@{}}
                \toprule
                    \multicolumn{1}{c}{\multirow{2}{*}{}} & \multicolumn{2}{c}{$\NumThreshold=50$} & \multicolumn{2}{c}{$\NumThreshold=100$} & \multicolumn{2}{c}{$\NumThreshold=150$}  & \multicolumn{2}{c}{$\NumThreshold=200$}  & \multicolumn{2}{c}{$\NumThreshold=250$}  \\
                       & \multicolumn{1}{c}{\Accuracy \%} & \multicolumn{1}{c}{$\PValue < 0.05$} & \multicolumn{1}{c}{\Accuracy \%} & \multicolumn{1}{c}{$\PValue < 0.05$} & \multicolumn{1}{c}{\Accuracy \%} & \multicolumn{1}{c}{$\PValue < 0.05$}  & \multicolumn{1}{c}{\Accuracy \%} & \multicolumn{1}{c}{$\PValue < 0.05$}  & \multicolumn{1}{c}{\Accuracy \%} & \multicolumn{1}{c}{$\PValue < 0.05$}  \\
                \midrule
                \ReferenceMethod & $27.82(\pm1.50)$ & $3/20$ & $32.48(\pm1.39)$ & $4/20$ & $43.66(\pm1.14)$ & $6/20$ & $47.74(\pm1.09)$ & $6/20$ & $52.19(\pm0.56)$ & $8/20$  \\
               \RandomNoise & $\mathbf{56.73(\pm0.57)}$ & $10/20$ & $\mathbf{57.14(\pm0.74)}$ & $10/20$ & $\mathbf{57.51(\pm0.91)}$ & $14/20$ & $\mathbf{58.79(\pm0.84)}$ & $15/20$ & $\mathbf{58.85(\pm0.81)}$ & $14/20$  \\
                \FriendlyNoise & $\mathbf{56.36(\pm0.95)}$ & $9/20$ & $\mathbf{57.45(\pm0.68)}$ & $14/20$ & $\mathbf{57.60(\pm0.71)}$ & $14/20$ & $\mathbf{58.64(\pm0.78)}$ & $16/20$ & $\mathbf{58.44(\pm0.60)}$ & $16/20$  \\
                \sysName & $\mathbf{56.65(\pm1.17)}$ & $\mathbf{8/20}$ & $\mathbf{57.65(\pm0.80)}$ & $\mathbf{9/20}$ & $\mathbf{57.89(\pm0.79)}$ & $\mathbf{8/20}$ & $\mathbf{58.59(\pm0.82)}$ & $\mathbf{8/20}$ & $\mathbf{58.90(\pm0.83)}$ & $\mathbf{10/20}$  \\
                \bottomrule
                \end{tabular}
                }}
            
                \caption{Experimental results of purification against \emph{data tracing}
                  on TinyImageNet where
                  \knnParameter-NN(\NumThreshold) detectors were applied. The
                  precision and recall of such detectors are shown in
                  \tblref{table:tracing:knn-kappa:tinyimagenet:parameters} in
                  \appref{sec:app:knn}. We bold the numbers following the same
                  rule as \tblref{table:poisoning:human:cifar10}.}
                \label{table:tracing:knn-kappa:tiny}
            \end{table*}

\section{Adaptive Training Data Manipulation}
\label{app:adaptive}

We considered a setting where the data contributor knows the deployment of \sysName. 
As such, the data contributor applies adaptive training data manipulation strategies by
tuning the perturbation bound (\PoisonHyperparameter). Specifically,
the data contributor tunes down \PoisonHyperparameter such that the manipulated data is similar
to the unmanipulated one to degrade the performance of \sysName. In this study, we investigate the effectiveness
of \sysName on CIFAR-10 and CIFAR-100 when \PoisonHyperparameter of the manipulation is changed from $16$ to $4$ for data poisoning
and from $10$ to $2$ for data tracing. To implement \sysName, we used a simulated detector with $\Precision=1.0$ and
$\Recall=0.1$ to select reference data, and set the same hyperparameters as those used in \secref{sec:poisoning} and \secref{sec:tracing}.  

The experimental results from adaptive manipulation techniques are shown 
in \tblref{table:adaptive}, where we compared the performances of \sysName 
with those from no purification applied. When \PoisonHyperparameter is tuned down from $16$ to $8$
in adaptive data poisoning, our proposed method is degraded slightly by the adaptive strategies
but still effective to defend against the attacks. When \PoisonHyperparameter continues to decrease to
$4$, the effectiveness of the attacks is diminished and thus our proposed method can easily
remove the manipulation from data poisoning. We have similar observations from the results
of adaptive data tracing. These results demonstrate the robustness of \sysName to the adaptive
strategies.

\begin{table*} [ht!]
\centering
\vspace{0.1in}
{ \resizebox{0.95\textwidth}{!}{
    \begin{tabular}{@{}lr@{\hspace{1.5em}}rr@{\hspace{1.5em}}rr@{\hspace{1.5em}}rr@{\hspace{1.5em}}rr@{\hspace{1.5em}}r@{}}
        \toprule
        \multicolumn{1}{c}{\multirow{2}{*}{}} & \multicolumn{2}{c}{$\PoisonHyperparameter=16$}& \multicolumn{2}{c}{$\PoisonHyperparameter=12$}& \multicolumn{2}{c}{$\PoisonHyperparameter=8$} & \multicolumn{2}{c}{$\PoisonHyperparameter=4$} \\
        & \multicolumn{1}{c}{\Accuracy \%} & \multicolumn{1}{c}{\AttackSuccessRate}& \multicolumn{1}{c}{\Accuracy \%} & \multicolumn{1}{c}{\AttackSuccessRate}& \multicolumn{1}{c}{\Accuracy \%} & \multicolumn{1}{c}{\AttackSuccessRate} & \multicolumn{1}{c}{\Accuracy \%} & \multicolumn{1}{c}{\AttackSuccessRate}  \\
        \midrule
        No purification & $91.97(\pm0.25)$ & $17/20$ & $92.08(\pm0.26)$ & $13/20$ &$92.20(\pm0.24)$ & $9/20$ & $92.15(\pm0.35)$ & $0/20$\\
        \sysName & $91.44(\pm0.42)$ & $0/20$ & $91.58(\pm0.44)$ & $0/20$ & $91.49(\pm0.36)$ & $2/20$ & $91.65(\pm0.50)$ & $0/20$\\
    \bottomrule
    \end{tabular}
    }}
    {\resizebox{0.95\textwidth}{!}{
        \begin{tabular}{@{}lr@{\hspace{1.5em}}rr@{\hspace{1.5em}}rr@{\hspace{1.5em}}rr@{\hspace{1.5em}}rr@{\hspace{1.5em}}r@{}}
            \toprule
            \multicolumn{1}{c}{\multirow{2}{*}{}} & \multicolumn{2}{c}{$\PoisonHyperparameter=16$} & \multicolumn{2}{c}{$\PoisonHyperparameter=12$} & \multicolumn{2}{c}{$\PoisonHyperparameter=8$} & \multicolumn{2}{c}{$\PoisonHyperparameter=4$} \\
            & \multicolumn{1}{c}{\Accuracy \%} & \multicolumn{1}{c}{\AttackSuccessRate \%} & \multicolumn{1}{c}{\Accuracy \%} & \multicolumn{1}{c}{\AttackSuccessRate \%} & \multicolumn{1}{c}{\Accuracy \%} & \multicolumn{1}{c}{\AttackSuccessRate \%} & \multicolumn{1}{c}{\Accuracy \%} & \multicolumn{1}{c}{\AttackSuccessRate \%}  \\
            \midrule
            No purification & $91.93(\pm0.39)$ & $89.47$ & $91.95(\pm0.30)$ & $59.54$ & $91.83(\pm0.28)$ & $36.01$ & $91.94(\pm0.26)$ & $6.97$\\
            \sysName & $91.30(\pm0.37)$ & $2.76$ & $91.07(\pm0.53)$ & $3.37$ & $91.37(\pm0.52)$ & $8.41$ & $91.26(\pm0.49)$ & $3.50$\\
        \bottomrule
        \end{tabular}
    }}
    {\resizebox{0.95\textwidth}{!}{
        \begin{tabular}{@{}lr@{\hspace{1.5em}}rr@{\hspace{1.5em}}rr@{\hspace{1.5em}}rr@{\hspace{1.5em}}rr@{\hspace{1.5em}}r@{}}
            \toprule
            \multicolumn{1}{c}{\multirow{2}{*}{}} & \multicolumn{2}{c}{$\TracingHyperparameter=10$} & \multicolumn{2}{c}{$\TracingHyperparameter=8$}& \multicolumn{2}{c}{$\TracingHyperparameter=5$} & \multicolumn{2}{c}{$\TracingHyperparameter=2$} \\
            & \multicolumn{1}{c}{\Accuracy \%} & \multicolumn{1}{c}{$\PValue<0.05$} & \multicolumn{1}{c}{\Accuracy \%} & \multicolumn{1}{c}{$\PValue<0.05$} & \multicolumn{1}{c}{\Accuracy \%} & \multicolumn{1}{c}{$\PValue<0.05$} & \multicolumn{1}{c}{\Accuracy \%} & \multicolumn{1}{c}{$\PValue<0.05$}  \\
            \midrule
            No purification & $70.98(\pm0.28)$ & $20/20$ & $71.24(\pm0.36)$ & $20/20$ & $71.76(\pm0.33)$ & $20/20$ & $72.64(\pm0.34)$ & $8/20$\\
            \sysName & $67.63(\pm0.28)$ & $4/20$ &$67.57(\pm0.26)$ & $6/20$ & $68.70(\pm0.38)$ & $5/20$ & $67.52(\pm0.36)$ & $0/20$\\
        \bottomrule
        \end{tabular}
    }}
        \caption{Experimental results of purification against \textit{adaptive} training data
        manipulation. The top table is for adaptive targeted data poisoning attacks on CIFAR-10; the middle
        table is for adaptive backdoor attacks on CIFAR-10; the bottom table is for adaptive data tracing on CIFAR-100. "No purification"
        refers to the setting where no purification was applied. 
        }
        \label{table:adaptive}
    \end{table*}


\end{document}